\documentclass[useAMS,usenatbib]{mnras}
\usepackage{savesym}
\usepackage{txfonts}
\savesymbol{iint}
\savesymbol{iiint}
\savesymbol{iiiint}
\savesymbol{idotsint}
\usepackage{graphicx}
\usepackage{amssymb}
\usepackage[below]{placeins}
\usepackage{txfonts}
\usepackage{natbib}
\usepackage{amsmath}
\usepackage{float}
\restoresymbol{TXF}{iint}
\restoresymbol{TXF}{iiint}
\restoresymbol{TXF}{iiiint}
\restoresymbol{TXF}{idotsint}
\usepackage{multirow}
\usepackage{array}
\usepackage{xcolor}
\usepackage[percent]{overpic}
\bibpunct{(}{)}{;}{a}{}{,}

\def \aap{A\&A}
\def \apjl{ApJ}
\def \pasj{PASJ}
\def \mnras{MNRAS}
\def \apj{ApJ}

\def \aj{AJ}
\def \apjs{ApJS}
\def\nat{Nature}

\def \xmm{{\emph{XMM-Newton}}}

\def \chandra{{\emph{Chandra}}}
\def \hitomi{{\emph{Hitomi}}}
\def \asca{{\emph{ASCA}}}

\def \suzaku{{\emph{Suzaku}}}

\def\physrep{Phys. Rep.}

\def\spose#1{\hbox to 0pt{#1\hss}}
\def\approxlt{\mathrel{\spose{\lower 3pt\hbox{$\sim$}}
        \raise 2.0pt\hbox{$<$}}}
\def\approxgt{\mathrel{\spose{\lower 3pt\hbox{$\sim$}}
        \raise 2.0pt\hbox{$>$}}}
\def\approxpropto{\mathrel{\spose{\lower 3pt\hbox{$\sim$}}
        \raise 2.0pt\hbox{$\propto$}}}
\mathchardef\twiddle="2218

\def\multleft#1{\hbox to size{\vbox {\halign {\lft{##}\cr #1}}\hfill}\par}
\def\multright#1{\hbox to size{\vbox {\halign {\rt{##}\cr #1}}\hfill}\par}

\def\today{\ifcase\month\or January\or February\or March\or April\or May\or
      June\or July\or August\or September\or October\or November\or December\fi
      \space\number\day, \number\year}
\def\<{\thinspace}

\def\arcsec{{\rm\thinspace arcsec}}

\def\arcsec {\hbox{$^{\prime\prime}$}}

\makeatletter
\newcommand{\thickhline}{%
    \noalign {\ifnum 0=`}\fi \hrule height 1.2pt
    \futurelet \reserved@a \@xhline
}
\newcolumntype{"}{@{\hskip\tabcolsep\vrule width 1pt\hskip\tabcolsep}}
\makeatother

\bibpunct{(}{)}{;}{a}{}{,}

\title[Chemical Enrichment History of the Perseus Cluster]{Constraints on the Chemical Enrichment History of the Perseus Cluster of Galaxies from High-Resolution X-ray Spectroscopy}

\author[Simionescu et al.]
{\parbox{\textwidth}{A. Simionescu$^{1,2,3}$, S. Nakashima$^{4}$, H. Yamaguchi$^{2,5,6}$, K. Matsushita$^7$, F. Mernier$^{8,9,1}$, N. Werner$^{8,10,11}$, T. Tamura$^{2}$, K. Nomoto$^{3}$, J. de Plaa$^{1}$, S.-C. Leung$^{3}$,
A. Bamba$^{12,13}$,
E. Bulbul$^{14}$, 
M. E. Eckart$^{5}$,
Y. Ezoe$^{15}$,
A. C. Fabian$^{16}$, 
Y. Fukazawa$^{11}$,
L. Gu$^{4}$,
Y. Ichinohe$^{17}$,
M. N. Ishigaki$^{3}$, 
J. S. Kaastra$^{1,18}$, 
C. Kilbourne$^5$,
T. Kitayama$^{19}$, 
M. Leutenegger$^{5}$,
M. Loewenstein$^{5,6}$, 
Y. Maeda$^1$, 
E. D. Miller$^{20}$, 
R. F. Mushotzky$^{6}$,
H. Noda$^{21}$,
C. Pinto$^{16}$,
F. S. Porter$^{5}$,
S. Safi-Harb$^{22}$,
K. Sato$^{23}$,
T. Takahashi$^{3}$,
S. Ueda$^{24,2}$,
S. Zha$^{25}$}\
\vspace{0.5cm}\\
\parbox{\textwidth}{
$^{1}$SRON Netherlands Institute for Space Research, Sorbonnelaan 2, 3584 CA Utrecht, The Netherlands\\
$^2$Institute of Space and Astronautical Science (ISAS), JAXA, 3-1-1 Yoshinodai, Chuo-ku, Sagamihara, Kanagawa, 252-5210, Japan\\
$^{3}$Kavli Institute for the Physics and Mathematics of the Universe (WPI), The University of Tokyo, Kashiwa, Chiba 277-8583, Japan \\
$^4$RIKEN High Energy Astrophysics Laboratory, 2-1 Hirosawa, Wako, Saitama 351-0198, Japan\\
$^5$NASA, Goddard Space Flight Center, 8800 Greenbelt Road, Greenbelt, MD 20771, USA \\
$^6$Department of Astronomy, University of Maryland, College Park, MD 20742, USA \\
$^7$Department of Physics, Tokyo University of Science, 1-3 Kagurazaka, Shinjuku-ku, Tokyo 162-8601, Japan \\
$^{8}$MTA-E\"otv\"os Lor\'and University Lend\"ulet Hot Universe Research Group, H-1117 P\'azm\'any P\'eter s\'et\'any 1/A, Budapest, Hungary\\
$^9$Institute of Physics, E\"otv\"os University, P\'azm\'any P\'eter s\'et\'any 1/A, Budapest, 1117, Hungary\\
$^{10}$Department of Theoretical Physics and Astrophysics, Faculty of Science, Masaryk University, Kotlarsk\'a 2, 611 37 Brno, Czech Republic\\
$^{11}$School of Science, Hiroshima University, 1-3-1 Kagamiyama, Higashi-Hiroshima 739-8526, Japan\\
$^{12}$Department of Physics, The University of Tokyo, 7-3-1 Hongo, Bunkyo-ku, Tokyo 113-0033, Japan \\
$^{13}$Research Center for the Early Universe, School of Science, The University of Tokyo, 7-3-1 Hongo, Bunkyo-ku, Tokyo 113-0033, Japan \\
$^{14}$Harvard-Smithsonian Center for Astrophysics, 60 Garden street, Cambridge 02138, Massachusetts, USA\\
$^{15}$Department of Physics, Tokyo Metropolitan University, 1-1 Minami-Osawa, Hachioji, Tokyo, 192-0397, Japan \\
$^{16}$Institute of Astronomy, University of Cambridge, Madingley Road, Cambridge, CB3 0HA, UK \\
$^{17}$Department of Physics, Rikkyo University, 3-34-1 Nishi-Ikebukuro, Toshima-ku, Tokyo 171-8501, Japan\\
$^{18}$Leiden Observatory, Leiden University, PO Box 9513, 2300 RA Leiden, The Netherlands\\
$^{19}$Department of Physics, Toho University, 2-2-1 Miyama, Funabashi, Chiba 274-8510, Japan\\
$^{20}$Kavli Institute for Astrophysics and Space Research, Massachusetts Institute of Technology, 77 Massachusetts Avenue, Cambridge, MA 02139, USA\\
$^{21}$Department of Earth and Space Science, Osaka University, 1-1 Machikaneyama-cho, Toyonaka, Osaka 560-0043, Japan\\
$^{22}$Department of Physics and Astronomy, University of Manitoba, Winnipeg, MB R3T 2N2, Canada\\
$^{23}$Department of Physics, Saitama University, 255 Shimo-Okubo, Sakura-ku, Saitama, 338-8570, Japan\\
$^{24}$Academia Sinica Institute of Astronomy and Astrophysics (ASIAA), P.O. Box 23-141, Taipei 10617, Taiwan\\
$^{25}$Physics Department, The Chinese University of Hong Kong, Hong Kong S.A.R., China \\
}}

\begin{document}

\maketitle

\begin{abstract}

High-resolution spectroscopy of the core of the Perseus Cluster of galaxies, using the \hitomi\ satellite above 2~keV and the \xmm\ Reflection Grating Spectrometer at lower energies, provides reliable constraints on the abundances of O, Ne, Mg, Si, S, Ar, Ca, Cr, Mn, Fe, and Ni. Accounting for all known systematic uncertainties, the Ar/Fe, Ca/Fe, and Ni/Fe ratios are determined with a remarkable precision of less than 10\%, while the constraints on Si/Fe, S/Fe, and Cr/Fe are at the 15\% level, and Mn/Fe is measured with a 20\% uncertainty. The average biases in determining the chemical composition using archival CCD spectra from \xmm\ and \suzaku\ range typically from 15--40\%. 
A simple model in which the enrichment pattern in the Perseus Cluster core and the proto-solar nebula are identical gives a surprisingly good description of the high-resolution X-ray spectroscopy results, with $\chi^2=10.7$ for 10 d.o.f. However, this pattern is challenging to reproduce with linear combinations of existing supernova nucleosynthesis calculations, particularly given the precise measurements of intermediate $\alpha$-elements enabled by \hitomi. 
We discuss in detail the degeneracies between various supernova progenitor models and explosion mechanisms, and the remaining uncertainties in these theoretical models. We suggest that including neutrino physics in the core-collapse supernova yield calculations may improve the agreement with the observed pattern of $\alpha$-elements in the Perseus Cluster core.
Our results provide a complementary benchmark for testing future nucleosynthesis calculations required to understand the origin of chemical elements. 

\end{abstract}


\section{Introduction}\label{intro}

Shortly after the Big Bang, all ordinary matter in the Universe was in the form of hydrogen and helium gas, with only trace amounts of other light elements like Li and Be. The chemical elements required for the existence of life were later forged in stars and dispersed into the interstellar medium by supernova explosions \citep{burbidge1957}. 
Only a small fraction of the original gas reservoir has undergone this process: 70 -- 90 \% of the baryonic mass content in clusters of galaxies are in the form of hot ($10^7-10^8$ K), diffuse X-ray emitting gas \citep{ettori1999} that fills the space between galaxies (so-called intra-cluster medium, or ICM). Nevertheless, metals produced in cluster member galaxies have been expelled into the ICM, such that a dominant fraction of the chemical elements in the local Universe is now found in this X-ray emitting plasma.

Remarkably, the spectral band of modern X-ray telescopes covers emission lines from all the abundant chemical elements (from C up to Ni; in astronomy, these are all collectively referred to as ``metals''). Moreover, interpreting spectra of the ICM is relatively uncomplicated, because the plasma is (with very few exceptions) optically thin and in collisional ionisation equilibrium. This makes X-ray spectroscopy an ideal tool to probe the chemical evolution history of the Universe.

The ICM in clusters of galaxies serves as the repository for ejecta from billions of type Ia and core-collapse supernova explosions (SNIa and SNcc, respectively), with very different abundance ratio patterns. Broadly speaking, the lighter metals, from O to Mg, are mainly produced in massive stars and ejected in SNcc at the end of their life time, while heavier elements from Cr to Ni are mainly produced by SNIa \citep[see, e.g.][]{tsujimoto1995}. Both supernova types contribute significant amounts of intermediate elements from Si to Ca.
The fraction of SNIa to SNcc enriching the ICM is therefore one of the main factors that affects the observed metal abundance patterns. The initial metallicity of the progenitors, the initial mass function (IMF) of the stars that explode as SNcc, and details about the exact SNIa explosion mechanism, also influence the observed ratios between various elements (see reviews by \citealt{werner2008} and \citealt{mernier_rev}).

Abundance measurements from \asca, and later \xmm\ and \suzaku\ observations, have been used by several groups to calculate the contribution of different kinds of supernovae to the ICM enrichment and to put constraints on the theoretical supernova models and properties of the stellar population responsible for the chemical enrichment of the Universe on the largest scales \citep[e.g.][]{mushotzky1996,deplaa2007, sato2007}. In what is arguably the most comprehensive study on this topic to date, \citet{mernier2016i} used a large sample of \xmm\  observations to constrain the average abundances of 11 different metals (O, Ne, Mg, Si, S, Ar, Ca, Cr, Mn, Fe, Ni) in the central ICM of 44 clusters of galaxies. Based on those measurements, \citet{mernier2016ii} discuss the results and limitations of trying to fit the measured chemical enrichment pattern with various combinations of nucleosynthesis models from the literature. This study estimates that the fraction of SNIa over the total number of SNe (SNIa+SNcc) contributing to the ICM enrichment ranges between 29--45\%, in good agreement with previous estimates \citep[e.g.][]{simionescu2009a,bulbul2012}. However, \citet{mernier2016ii} also show that combinations of any commonly employed SNIa and SNcc yields fail to reproduce the abundance pattern in detail, with the observed Ca/Fe and Ni/Fe ratios in particular being significantly underestimated by the models.

Some of this tension has been alleviated by the recent high-spectral resolution observations of the Perseus Cluster performed with the Soft X-ray-Spectrometer (SXS) onboard the \hitomi\ satellite \citep{takahashi2016}. The superb energy resolution of the SXS of only 5~eV allows us to separate emission lines from Ni and Fe, which appear blended in spectra measured with conventional CCDs, and to test underlying atomic models. The resulting best-fit Ni/Fe ratio is much lower than initially reported in \citet{mernier2016i}. In addition, weak lines from Cr and Mn are seen with a much better contrast against the continuum emission. \citet{HitomiNatureZ} (hereafter ZNpaper) focusses on the Cr/Fe, Mn/Fe, and Ni/Fe ratios measured accurately for the first time with the SXS using an integrated spectrum from the entire field of view of the \hitomi\ observation, and discusses their implications for constraining the SN~Ia progenitors.  

On the other hand, the $\alpha$-element to Fe ratios, together with the absolute Fe/H, are also a powerful diagnostic both of supernova nucleosynthesis details and of the stellar evolution in the cluster member galaxies responsible for polluting the ICM with metals. Early X-ray data had suggested that the $\alpha$-element to Fe abundance pattern was different from Solar \citep{mushotzky1996}, varied from system to system \citep[e.g.][]{fukazawa2000}, or changed as a function of distance from the brightest cluster galaxy (BCG) \citep[e.g.][]{finoguenov2000,rasmussen2009}. Given that very massive elliptical galaxies are known to have high $\alpha$-element to Fe ratios \citep{conroy2014}, while on the other hand the rate of SN~Ia per unit star formation may be higher in cluster galaxies than in the field \citep{maoz2017}, these differences in chemical composition between the ICM and the Milky Way were not surprising, and promised to yield important clues towards understanding how star formation proceeds differently as a function of environment. 

However, with improvements in the photon statistics, instrument calibration, and spectral modelling, more recent X-ray measurements are converging on a scenario wherein the abundance ratios in the ICM are in fact consistent with the Solar pattern. For the nearby, X-ray bright Perseus Cluster, near-Solar relative abundances with respect to Fe were found both in the core \citep{tamura2009} and on $\sim$~Mpc scales \citep{matsushita2013}; in-depth studies of other individual clusters show that the $\alpha$-element to Fe ratios in the ICM appear to be consistent with Solar over an order of magnitude in absolute metallicity, from the most metal-enriched cluster core (the Centaurus Cluster with a central Z=2 Solar, \citealt{sanders2006,matsushita2007}) all the way to the largest cluster-centric distances probed to date (0.2 Solar near the virial radius of the Virgo Cluster, \citealt{simionescu2015}). \citet{mernier2017} find that the relative abundances of all elements detected by \xmm\ with respect to Fe remain constant over a large sample of cluster cores, despite a radial gradient in [Fe/H].

These two emerging paradigms -- namely, a chemical composition of the ICM that is (i) close to Solar and (ii) spatially invariant -- need to be confirmed (or ruled out) by using the full capabilities of the best high-resolution spectroscopy observations that are accessible to date. In this paper, we expand upon the initial results of \hitomi\ to present the most detailed view so far of the chemical abundance pattern in the Perseus Cluster core. In particular, we derive constraints on the spatial variation of the measured metal abundance ratios as a function of radius, within the limits allowed by \hitomi's point spread function and field of view (Section \ref{sect_radial}); we show the results of searching for weak line emission from other elements such as Al, Cl, K, and Ti in the micro-calorimeter data (Section \ref{sect_rare}); we discuss the implications of a Solar abundance pattern in the Perseus Cluster core (Section \ref{sect_solar}); and we combine SXS and \xmm\ Reflection Grating Spectrometer (RGS) data to provide updated constraints on the supernova yields using the highest available spectral resolution for all elements from O through Ni that are detected in the X-ray band (Section \ref{sect_yields}), focusing in particular on the lighter $\alpha$-elements not previously discussed in the ZNpaper. 
In addition, the wealth of the Perseus observational data obtained with both CCD and high-resolution spectroscopy instruments provides an unique opportunity to accurately quantify the systematic biases in a consistent way. Therefore, we compare the results derived from high-resolution versus CCD spectroscopy in order to estimate the reliability of previous estimates of the chemical composition of the ICM (Section \ref{disc_sys}).

\section{Observations and Data Reduction}

We combine information from the recent \hitomi\ SXS observations, together with a reanalysis of archival data from \xmm\ and \suzaku, to provide the most detailed picture of the chemical composition of a galaxy cluster core available to date. The observations used in this work are listed in Table \ref{tab_obs}, and the data analysis methods for each instrument are described below. Due to different sensitivities, bandpasses, and spatial responses, different instruments will sample slightly different parts of the central ICM with different weightings, even when identical spatial extraction regions are considered. Therefore, in our analysis, the best-fit spectral model parameters are calculated separately for each of the various detectors considered here. Unless explicitly stated otherwise, statistical errors are quoted at the 68\% confidence level throughout the manuscript, and abundances are expressed in units of the proto-solar values reported by \citet{lodders2009}.

\begin{table}
\caption{List of observations used in this work}\label{tab_obs}
\vspace{0.5cm}
\centering
\begin{tabular}{ccc}
\hline
 Obs. Date & ObsID & Exposure \\
 & & time (ks) \\
\hline
\multicolumn{3}{c}{\hitomi}\\
\hline
 2016-02-24 & 10040010 & 49 \\
2016-02-25 & 10040020 & 97 \\
2016-03-04 & 100400[3,4,5]0 & 146 \\
2016-03-06 & 10040060 & 46 \\
\hline
\multicolumn{3}{c}{\xmm}\\ 
\hline
 2001-01-30 & 0085110[1,2]01 & 72 \\
 2006-01-29 & 0305780101 & 125 \\
\hline
\multicolumn{3}{c}{\suzaku}\\ 
\hline
2006-02-01 & 800010010 & 44 \\
2006-08-29 & 101012010 & 46 \\
2007-02-05 & 101012020 & 40 \\
2007-08-15 & 102011010 & 36 \\
2008-02-07 & 102012010 & 36 \\
2008-08-13 & 103004010 & 34 \\
2009-02-11 & 103004020 & 46 \\
2009-08-26 & 104018010 & 35 \\
2010-02-01 & 104019010 & 34 \\
2010-08-09 & 105009010 & 30 \\
2011-02-03 & 105009020 & 33 \\
2011-07-27 & 106005010 & 34 \\
2012-02-07 & 106005020 & 42 \\
2012-08-20 & 107005010 & 41 \\
2013-02-11 & 107005020 & 34 \\
2013-08-15 & 108005010 & 33 \\
2014-02-05 & 108005020 & 34 \\
2014-08-27 & 109005010 & 16 \\
2015-03-03 & 109005020 & 34 \\
\hline
\end{tabular}
\end{table}

\subsection{\hitomi\ Soft X-ray Spectrometer}

The data reduction procedure used in this work is identical to that presented in \citealt{HitomiT} (hereafter Tpaper). In summary, we use the cleaned event data in the third pipeline products, with the standard screening for the post-pipeline data reduction \citep{angelini2016}. The spectral analysis was performed using only GRADE Hp
(high-resolution primary) events that have the best energy resolution. The redistribution matrix (RMF) was generated with the XL (extra-large) size option, which accounts for the full components of the spectral response, including the main peak, low-energy exponential tail, escape peaks, and electron-loss continuum. The ancillary response file (ARF) was generated assuming a diffuse source with a radius of $12^\prime$ and with the surface brightness distribution constrained by a \chandra\ image extracted in the 1.8--9.0 keV band (excluding the AGN). The pixel-by-pixel redshift correction and parabolic gain correction described in the Tpaper were also applied (effectively meaning that all spectra are adjusted to match the redshift of 0.017284 determined in \citealt{HitomiV}). Non X-ray backgrounds were produced using the task \texttt{sxsnxbgen} and subtracted from the observed spectra.

Spectra were modelled in Xspec 12.9.1h \citep{arnaud1996}, employing the modified C-statistic \citep{cash1979}. Both AtomDB version 3.0.9 \citep{foster2012} and the SPEX Atomic Code and Tables (SPEXACT) version 3.03.00 \citep{kaastra1996} were used to calculate the plasma models under the assumption of collisional ionisation equilibrium (CIE). A \texttt{python} program was used to generate APEC format table models from the spectral fitting program SPEX 3\footnote{adapted from http://www.mpe.mpg.de/$\sim$jsanders/code/}, allowing us to perform a direct comparison of the results using a consistent treatment of all other assumptions and fit procedures.
Photoelectric absorption by cold matter in our Galaxy was modelled using the TBabs code version 2.3 \citep{wilms2000} with a fixed hydrogen column density of $1.38\times10^{21}$ cm$^{-2}$ \citep{kalberla2005}. For extraction regions that include the central AGN, we added to our model a power-law continuum and a neutral Fe K line with parameters fixed at the values described in \citet{HitomiAGN}. To account for the effects of resonant scattering described in \citet{HitomiRS}, the Fe XXV He-$\alpha$ w line was removed from the thermal plasma models and fitted separately with a gaussian component with free width and normalisation. 

We present the results obtained by modelling the cluster emission with both a single-temperature and a two-temperature model. In order to fully utilise the line resolving power of the SXS, as well as to avoid uncertainties related to the calibration of the effective area affecting the continuum level, for the single-temperature fit we allowed the line temperature and the continuum temperature to vary independently. This is implemented as the \texttt{bvvtapec} model in Xspec, and referred to as the `modified 1 CIE model' in the Tpaper and throughout the rest of this manuscript. For the two-temperature model (hereafter `2 CIE'), an additional component was added to the `modified 1 CIE model.' The line and continuum temperatures for the second component cannot be constrained independently, and were therefore assumed to have the same value. The metal abundances were linked between the two thermal components in the 2 CIE model.

\subsection{\xmm}

\subsubsection{Reflection Grating Spectrometer}

The Perseus Cluster is included in the CHemical Evolution RGS Sample (CHEERS) \citep{deplaa2017}, and we follow the data analysis methods laid out by the CHEERS collaboration. In brief, the RGS observations were reduced in the standard way using the \xmm\ Science Analysis Software (SAS) version 14.0.0. RGS light curves extracted from CCD number 9, where hardly any source emission is expected, were examined, and all time intervals with a count rate deviating from the mean by more than $2\sigma$ were excluded in order to avoid contamination from soft-proton flares.

We extract the RGS source spectra in a region centered on the peak of the source emission, with a width of $0.8^\prime$ in the cross-dispersion direction. We
use the model background spectrum created by the standard RGS pipeline, which is a template background file that is rescaled based on the count rate in CCD9 of the RGS. 
We use the SPEX software package to fit the RGS 1 and 2 source spectra for the observations listed in Table \ref{tab_obs} in parallel, with the relative instrument normalisations treated as free parameters to account for any small differences in calibration between the individual detectors and observations. 

Since the RGS is a slit-less spectrometer, photons originating from a region near the cluster center, but offset in the direction along the dispersion axis, will be slightly shifted in wavelength with respect to line emission from the cluster center. To correct for this effect, we use the surface brightness profile extracted from the \xmm\ MOS1 detector, whose DETY coordinate direction is parallel to the dispersion direction in RGS1 and RGS2. This spatial profile is convolved with the model spectrum during spectral fitting, using the \texttt{lpro} model component in SPEX.

The spectra were fit in the 8--21\AA\ wavelength range using a two-temperature model plus a power-law component accounting for the central AGN. The power-law index was fixed to 1.9 (with the normalization left free in the fit) and the redshift was fixed to 0.017284 (corresponding to the assumption for the \hitomi\ SXS analysis). The \texttt{lpro} model allows for wavelength shifts, including any uncertainty on the assumed redshift. Galactic absorption was modeled as a \texttt{hot} component in SPEX, with a fixed\footnote{when allowed as a free parameter in the 2T SPEXACT fit, the best-fit $n_{\rm H}$ is within 5\% of the assumed value from \citet{kalberla2005}.} hydrogen column density of $n_{\rm H}=1.38\times10^{21}$ cm$^{-2}$. The abundances of O, Ne, Mg, Fe, and Ni, as well as the spatial broadening and wavelength shift in the \texttt{lpro} model, were coupled between the two plasma temperature components and left free in the fit; the abundances of all other chemical elements are fixed to 1 Solar. The \texttt{lpro} model parameters were allowed to vary independently for observations performed in different years (2001 versus 2006). We model the spectra using both the AtomDB v3.0.9 and SPEXACT v3.03.00 emission line databases. However, because an equivalent of the \texttt{lpro} model is not available in Xspec, the AtomDB v3.0.9 fit is performed within SPEX. This is achieved by implementing a \texttt{user} model that calls Xspec externally for every function evaluation required by the fit and returns the model calculation to SPEX.

\subsubsection{European Photon Imaging Camera (EPIC)} \label{epic_analysis}

We reduced the data from the two MOS (metal oxide semiconductor) and one pn camera onboard \xmm\ in the standard way, using the Science Analysis System (SAS) v14.0.0. The procedure is described in detail in \cite{mernier2015} and summarised below. 

Soft proton flares are removed using light curves in the 10--12 keV energy band for MOS and 12--14 keV for pn, with a time binning of 100 s; time intervals for which the count rate deviates by more than $2\sigma$ from the mean value are excluded from the analysis. We then repeat the procedure for the 0.3--10 keV band using 10 s time bins.  
For spectroscopy, we impose the condition \texttt{FLAG==0} to select only the highest quality events, and we restrict the analysis to the single to quadruple pixel MOS events (\texttt{PATTERN $\leq$ 12}), and the single pixel pn events (\texttt{PATTERN==0}). 

We extract spectra from a $3\times3$ arcmin box centred on ($\alpha$ 3:19:44.56, $\delta$ +41:31:13.36), with a position angle of 73 degrees. This corresponds to the full field of view of the longest \hitomi\ SXS observations (ObsID 100400[2,3,4,5]0). A circular region with a radius of 10\arcsec\ around the central AGN was excised from the analysis. Instrument response files are generated using the dedicated tasks \texttt{rmfgen} and \texttt{arfgen}. 

We fit the spectra using the SPEX package, employing a two-temperature model modified by Galactic absorption, with the elemental abundances assumed to be the same for both thermal components. The metals whose abundances are not constrained by the fit are assumed to have the same concentration with respect to H (expressed in Solar units) as Fe. The Galactic absorption is described as a \texttt{`hot'} model component with $kT=0.5$ eV. We model the local hot bubble, galactic thermal emission, unresolved point sources, hard particle background, and soft-proton background following the method described extensively in \citet{mernier2015,mernier2016i}. The temperatures, spectral normalisations, and abundances of the model components describing the cluster emission are allowed to vary independently for MOS and pn. The Galactic absorption column density was free to vary, but assumed to be the same for all spectra; as a consequence, the O abundance was also assumed to be the same for both MOS and pn. To counteract small differences in calibration, the redshifts and an overall normalisation factor of order unity were allowed as free parameters for each detector of each observation. Since Ne lines are blended within the Fe-L complex, and since the shape of this spectral feature provides a strong constraint on the multi-temperature structure, we have chosen to fix the Ne/Fe value in the fit to that determined using the RGS. 

The parameters (temperature, metal abundances, and spectral normalisation) of the two CIE models were first constrained based on a `global fit' in the 0.6--10 keV band for pn and 0.5--10 keV for MOS. For all elements except Ne and Fe (for which Fe L-shell emission is important in the spectrum), we also re-fit EPIC spectra within local bands successively centred around the strongest K-shell lines of each element, leaving only the (local) normalisation, redshift, and the abundance of that metal free and fixing all other parameters to their global best-fit values (hereafter referred to as `local fits'). The boundaries of these local energy bands are, respectively, 0.51--0.77 keV for O, 1.36--1.59 keV for Mg, 1.65--2.26 keV for Si, 2.37--2.77 keV for S, 2.77--3.50 keV for Ar, 3.60--4.44 keV for Ca, 5.01--5.90 keV for Cr, 5.70--6.29 keV for Mn, and 7.24--8.02 keV for Ni.

\subsection{\suzaku\ X-ray Imaging Spectrometer} \label{xis_analysis}
The Perseus Cluster was used as a \suzaku\ calibration source and, as such, was observed regularly throughout the lifetime of the satellite.
We analyzed the data from all observations in the \suzaku\ archive where the front illuminated XIS 0, 2 and 3, and back illuminated XIS 1, were used in normal clocking mode and with no window option. 
XIS 2 was lost to a putative micrometeoroid hit on 2006 November 9 and therefore its data are available only for a small fraction of the observations. 
The total exposure time with \suzaku\ XIS is $682$ ks.

Initial cleaned event lists were obtained using the standard screening criteria proposed by the XIS team \footnote{Arida, M., XIS Data Analysis, http://heasarc.gsfc.nasa.gov/docs/suzaku/ analysis/abc/node9.html.}. We filtered out times of low geomagnetic cut-off rigidity (imposing the condition COR $>$ 6 GV). For the XIS 1 data obtained after the reported charge-injection level increase on 2011 June 1, we have excluded two adjacent rows on either side of the charge-injected rows. The instrumental background was determined in a standard way using night-Earth data. \citet{tamura2009} estimate that the cosmic X-ray background is well below 1\% of the source over almost the entire energy band, and we have therefore ignored this cosmic background in the analysis. 

As for the EPIC analysis, we extract spectra from a spatial region corresponding to the full field of view of the longest \hitomi\ SXS observations. However, the \suzaku\ point spread function does not allow us to isolate the emission from the central AGN; hence, this is modelled as a separate spectral component.
We fit the data in XSPEC using the 0.6--8.5 keV energy band for the back-illuminated XIS1, and 0.7--8.5 keV for the front-illuminated XIS0,2,3. To avoid calibration errors, we ignored energy ranges around the Si-K edge (1.6--1.95 keV) and the Au-M edge (2.2--2.4 keV). We model the spectra using a two-temperature model plus a power-law component with free normalization and a fixed index of 1.9. We present results using both AtomDB 3.0.9 and SPEXACT 3.03.00. The Galactic absorption column density and redshift were left as free parameters in the fit. The data for each detector of each observation was fit with the same spectral model, up to an overall normalization constant of order 1 that was left free in the fit in order to mitigate the effects of small differences in calibration between the individual detectors and their possible evolution over the lifetime of the satellite. \suzaku\ XIS data allows us to constrain the abundances of O, Ne, Mg, Si, S, Ar, Ca, Cr, Mn, Fe, and Ni, which were assumed to be the same for both plasma temperature components. 

In order to allow a closer comparison to the \xmm\ EPIC analysis, we also present the constraints obtained by fixing the SPEXACT model parameters to their values measured from the broad band, and re-fitting the \suzaku\ spectra within narrow energy ranges centred around the strongest K-shell lines of each element, allowing the abundance of that respective element to vary. For these `local fits', an overall normalisation of order unity was left free for each detector of each observation, allowing the local continuum to adjust in order to compensate for any calibration inaccuracies; one overall redshift (coupled between all \suzaku\ spectra) was also free to vary during the fit procedure.

\section{Results}


\subsection{Constraints on the spatial variation of metal abundance ratios obtained with the Hitomi SXS}\label{sect_radial}

\begin{table*}
\caption{Best-fit abundance ratios measured using the Hitomi SXS data for the different spatial regions considered in this work; the region choice is identical to that described in the Tpaper (see Figure 1 of that manuscript for details). Systematic differences due to the choice of effective area calibration are also shown for the highest statistical quality `entire core' region.}
\label{hitomiztab}
\centering
\footnotesize
\begin{tabular}{ccccccccc}
\hline
region/model           & Fe                & Si/Fe                  & S/Fe                   & Ar/Fe                  & Ca/Fe                  & Cr/Fe                  & Mn/Fe                  & Ni/Fe                  \\
\hline
\hline\multicolumn{1}{l}{Entire core}     \\
\multicolumn{1}{l}{~~~APEC:modified-1CIE} & $0.76\pm0.01$ & $0.88\pm0.05$ & $0.99\pm0.03$ & $0.91\pm0.04$ & $0.97\pm0.03$ & $0.84\pm0.11$ & $0.89\pm0.16$ & $0.91\pm0.05$ \\
\multicolumn{1}{l}{~~~APEC:2CIE}   & $0.81\pm0.01$  & $0.75\pm0.04$ & $0.86\pm0.02$ & $0.84\pm0.04$ & $0.92\pm0.03$ & $0.86\pm0.11$ & $0.92\pm0.16$ & $0.92\pm0.06$ \\
\multicolumn{1}{l}{~~~SPEX:modified-1CIE} & $0.78\pm0.01$ & $1.00\pm0.05$ & $1.08\pm0.03$ & $0.97\pm0.04$ & $1.05\pm0.04$ & $0.89\pm0.11$ & $0.98\pm0.17$ & $0.89\pm0.05$ \\
\multicolumn{1}{l}{~~~SPEX:2CIE}      & $0.86\pm0.01$ & $0.77\pm0.04$ & $0.84\pm0.02$ & $0.82\pm0.04$ & $0.96\pm0.03$ & $0.92\pm0.11$ & $1.03\pm0.18$ & $0.90\pm0.06$ \\
\hline\multicolumn{1}{l}{Nebula}          \\
\multicolumn{1}{l}{~~~APEC:modified-1CIE} & $0.77\pm0.01$ & $0.91\pm0.07$ & $1.03\pm0.04$ & $0.96\pm0.05$ & $0.98\pm0.04$ & $0.74\pm0.14$ & $0.99\pm0.21$ & $1.00\pm0.07$ \\
\multicolumn{1}{l}{~~~APEC:2CIE}      & $0.82\pm0.02$ & $0.77\pm0.06$ & $0.89\pm0.04$ & $0.87\pm0.05$ & $0.93\pm0.04$ & $0.75\pm0.14$ & $1.01\pm0.21$ & $1.01\pm0.08$ \\
\multicolumn{1}{l}{~~~SPEX:modified-1CIE} & $0.78\pm0.02$ & $1.04\pm0.07$ & $1.13\pm0.04$ & $1.02\pm0.06$ & $1.07\pm0.05$ & $0.78\pm0.14$ & $1.09\pm0.23$ & $0.98\pm0.07$ \\
\multicolumn{1}{l}{~~~SPEX:2CIE}       & $0.89\pm0.02$ & $0.78\pm0.07$ & $0.87\pm0.04$ & $0.86\pm0.05$ & $0.98\pm0.06$ & $0.80\pm0.15$ & $1.13\pm0.24$ & $0.99\pm0.08$ \\
\hline\multicolumn{1}{l}{Rim}             \\
\multicolumn{1}{l}{~~~APEC:modified-1CIE} &  $0.72\pm0.01$ & $0.83\pm0.08$ & $0.92\pm0.05$ & $0.85\pm0.06$ & $0.97\pm0.06$ & $1.07\pm0.17$ & $0.76\pm0.23$ & $0.74\pm0.08$ \\
\multicolumn{1}{l}{~~~APEC:2CIE}         & $0.73\pm0.01$ & $0.77\pm0.08$ & $0.86\pm0.04$ & $0.80\pm0.06$ & $0.93\pm0.05$ & $1.09\pm0.17$ & $0.79\pm0.24$ & $0.75\pm0.08$ \\
\multicolumn{1}{l}{~~~SPEX:modified-1CIE} & $0.74\pm0.01$  & $0.94\pm0.09$ & $1.01\pm0.05$ & $0.91\pm0.07$ & $1.05\pm0.06$ & $1.11\pm0.17$ & $0.82\pm0.26$ & $0.72\pm0.08$ \\
\multicolumn{1}{l}{~~~SPEX:2CIE}         & $0.79\pm0.01$ &  $0.77\pm0.07$ & $0.84\pm0.04$ & $0.79\pm0.06$ & $0.97\pm0.06$ & $1.13\pm0.17$ & $0.86\pm0.27$ & $0.74\pm0.08$ \\
\hline\multicolumn{1}{l}{Outer}           \\
\multicolumn{1}{l}{~~~APEC:modified-1CIE}  & $0.57\pm0.03$ & $0.89_{-0.41}^{+0.36}$ & $0.78\pm0.20$ & $0.91\pm0.29$ & $1.24\pm0.27$ & $1.54\pm0.66$ & $0.41_{-0.86}^{+0.41}$ & $0.91\pm0.27$ \\
\multicolumn{1}{l}{~~~SPEX:modified-1CIE} &  $0.59\pm0.03$ &  $0.99_{-0.42}^{+0.36}$ & $0.85\pm0.20$ & $0.97\pm0.30$ & $1.30\pm0.28$ & $1.42\pm0.66$ & $0.37_{-0.89}^{+0.37}$ & $0.88\pm0.25$ \\
\hline
\hline\multicolumn{2}{l}{Entire core: ARF systematics}     \\
\multicolumn{1}{l}{~~~APEC:2CIE:ground}   & $0.77\pm0.01$  & $0.90\pm0.05$ & $0.97\pm0.03$ & $0.86\pm0.04$ & $0.87\pm0.03$ & $0.80\pm0.10$ & $0.97\pm0.15$ & $0.99\pm0.06$ \\
\multicolumn{1}{l}{~~~SPEX:2CIE:ground}      & $0.84\pm0.01$ & $0.89\pm0.05$ & $0.93\pm0.03$ & $0.83\pm0.04$ & $0.89\pm0.03$ & $0.85\pm0.10$ & $1.06\pm0.18$ & $1.03\pm0.06$ \\
\multicolumn{1}{l}{~~~APEC:2CIE:crab}   & $0.87\pm0.01$  & $0.75\pm0.04$ & $0.84\pm0.02$ & $0.80\pm0.03$ & $0.89\pm0.03$ & $0.82\pm0.11$ & $0.86\pm0.16$ & $0.97\pm0.06$ \\
\multicolumn{1}{l}{~~~SPEX:2CIE:crab}      & $0.95\pm0.01$ & $0.75\pm0.04$ & $0.81\pm0.02$ & $0.78\pm0.03$ & $0.92\pm0.03$ & $0.86\pm0.11$ & $0.97\pm0.18$ & $0.96\pm0.06$ \\

\hline
\end{tabular}
\end{table*}

As detailed in the Tpaper, we use the Hitomi SXS observations to investigate the spectral properties for four different spatial regions. The `entire core' refers to the combined spectra from ObsIDs 10040020, 10040030, 10040040, 10040050, and 10040060. To look for any spatial variations, we divided this `entire core' region into the `nebula' region, where H$\alpha$ emission was reported by \citet{conselice2001}, and the `rim' region located just beyond this multi-phase core. Lastly the fourth region, which we refer to as the `outer region', is the entire FoV of ObsID 10040010, which is offset by $\sim3^\prime$ towards the west-southwest of the Perseus Cluster core. We refer the reader to Figure 1 of the Tpaper for a visual representation of the choice of spatial regions, and to the contents of that manuscript for a discussion of the detailed thermal structure obtained from this analysis. 

The best-fit Fe abundance and ratios of other elements with respect to Fe are summarised in Table \ref{hitomiztab}. For the `entire core', `nebula', and `rim' regions, we show the results using both the `modified 1 CIE' and `2 CIE' models described above; in the `outer' region, including a second temperature component does not improve the fit significantly and hence we only show the metal abundances obtained from a single-temperature (`modified 1 CIE') model. We find an encouraging agreement between the results obtained with AtomDB and SPEXACT. On the other hand, the best-fit metal abundances are sensitive to the assumed temperature structure, with the `2 CIE' model giving lower $\alpha$-element to Fe ratios than the `modified 1 CIE' model. 
As discussed in the Tpaper, the effective area calibration is an additional source of uncertainty that affects the measurements. Since the in-flight calibration of Hitomi was not completed because of its short life time, we assessed this uncertainty using the modified ancillary response file (ARF) based on the ground telescope calibration and the observed Crab data \citep{HitomiCrabARF}. The best-fit metal abundance ratios obtained for the highest quality `entire core' spectrum with these alternative choices of effective area calibration are also given in Table \ref{hitomiztab}; the magnitude of the systematic error is similar to the extent of the differences between the `modified 1 CIE' and `2 CIE' models.

\begin{figure}
\begin{center}
\includegraphics[width=\columnwidth]{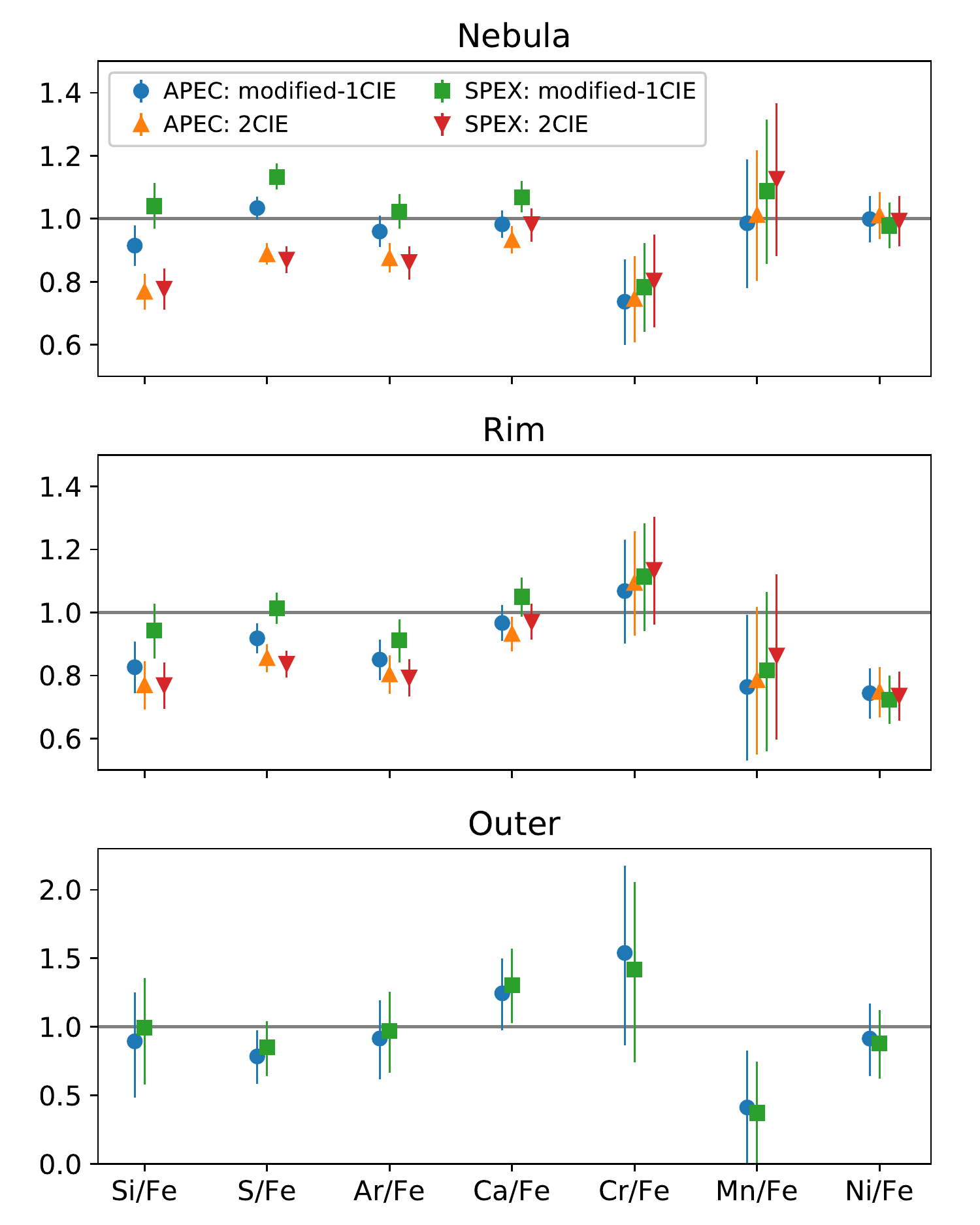} 
\end{center}
\caption{Abundance ratios with respect to Fe measured from the \hitomi\ SXS data using the three distinct spatial sub-regions considered here\textsuperscript{a}.}\label{radial}
\small\textsuperscript{a} Here, and in other table and figure captions throughout the text where appearing next to APEC, `SPEX' is exceptionally used as shorthand for `SPEXACT', in the interest of brevity. 
\end{figure}

Figure \ref{radial} compares the measured chemical abundance patterns for the three spatially distinct regions considered here (`nebula', `rim', and `outer'), using the standard ARF. While we find a significant radial gradient in the Fe abundance (that decreases by $\sim30$\% from the `nebula' to the `outer' region), taking into account the systematic uncertainties described above, no significant radial trends in the relative abundances of other chemical elements with respect to Fe are identified; the chemical composition of the ICM is roughly consistent with the Solar metal abundance pattern in all of the spatial regions considered.

\subsection{Search for emission lines from rare elements}\label{sect_rare}

\begin{figure}
\begin{center}
\includegraphics[width=\columnwidth]{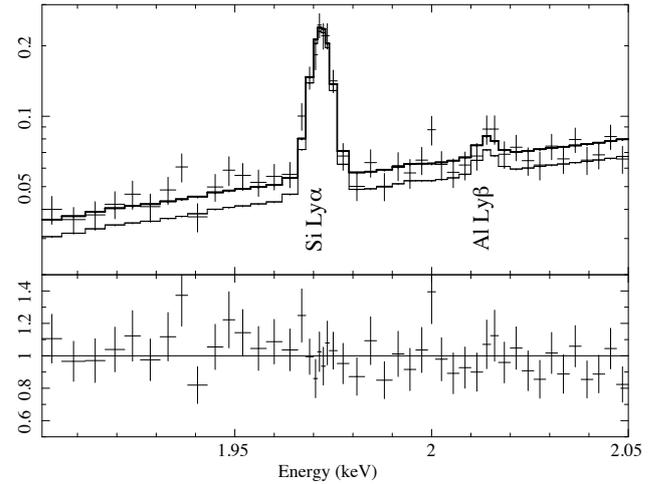} 
\end{center}
\caption{The `entire core' spectrum in the energy band including Si and Al transitions.
The observed flux in counts s$^{-1}$ keV$^{-1}$ and data to model ratio from the local fit described in the text are shown in the top and bottom panels.
In the top panel, the total emission and the ICM AtomDB component are shown by thick and thin histogram lines, respectively.}\label{al_localfit}
\end{figure}

One of the main advantages of high-resolution spectroscopy is the increased sensitivity to weak emission lines, and accuracy with which these lines can be separated from the underlying continuum emission. The \hitomi\ SXS observations of the core of the Perseus Cluster allow us to easily identify emission lines from Cr and Mn (see also ZNpaper). Hence, we have also searched for emission from even weaker lines of other rare elements using the spectra extracted from the `entire core' region. No detection is found with high confidence, although hints of Al Ly${\beta}$, a blend of Cl Ly${\alpha}$ and K He${\alpha}$, and Ti He${\alpha}$ are seen with a significance between $1-2\sigma$. This is expected, as \citet{kitayama2014} report that an exposure time of at least 625~ks would be required to detect P, K, Ti, and Co with a significance above $5\sigma$; Al and Cl should be detected in much shorter exposure times using the full capabilities of the SXS, but are not seen in our observation because the closed gate valve limits the band pass and effective area at softer energies. 

In order to obtain approximate constraints on the abundances of Al, Cl, K, and Ti, we then fitted the `modified 1 CIE' model to the entire core spectrum, restricting the fit to the local energy band that covers the brightest emission lines from the corresponding rare element in each case (1.9--2.05 keV for Al, 2.65--3.03 keV for Cl, 3.3--3.7 keV for K, and 5.44--4.72 keV for Ti). All model parameters were fixed based on the Tpaper, except the overall spectrum normalisation (allowed to vary in order to account for any uncertainties in calibrating the local continuum), the redshift, and the abundances of any elements whose emission lines are found in the considered local energy band. Figure \ref{al_localfit} shows an example of the spectral fit around the Al Ly${\beta}$ emission line (at a rest-frame energy of 2048.1 eV). For spectral plots of other energy bands containing the Cl, K, and Ti lines, we refer the reader to Figure 1 of \mbox{\cite{Hitomi35keV}} and Figure 24 of \citet{HitomiAtomic} (hereafter Atomic paper). 

Using AtomDB 3.0.9, we obtain $1\sigma$ confidence intervals of (2.2--7.8) for Al/Fe, (0.9--2.8) for Cl/Fe, (0.1--1.3) for K/Fe, and (0.4--1.6) for Ti/Fe. Given the very large statistical uncertainties, systematic errors due to the assumed temperature structure (`modified 1 CIE' or `2 CIE'), as well as the atomic line emission models (AtomDB or SPEXACT) are subdominant and not reported here. 

Within these large statistical uncertainties, the abundances of the rare elements with respect to Fe appear consistent with the Solar abundance pattern. We note, however, low-significance hints of super-solar Al/Fe. Nucleosynthesis models indicate that the abundances of odd-Z elements like Al and Na depend on the initial metallicity of core-collapse supernova progenitors \citep{nomoto2006, kobayashi2006, nomoto2013}. While the current data quality does not allow us to draw any firm conclusions, a high abundance of Al, if confirmed by future measurements, would be a particularly interesting tracer of the properties of the underlying stellar population responsible for the enrichment. An alternative explanation for the inferred high abundance of Al is that its Ly${\beta}$ line may be blended with satellite transitions of Fe XXV charge exchange (CX) emission around a rest frame of 2.05~keV \citep{gu2016}. The Atomic paper suggests a hint for Fe XXV CX direct emission at $\sim 8.6$~keV at $2.4\sigma$ significance, in which case weak emission from additional satellite lines could be naturally expected.

\subsection{Abundances of light elements measured from the \xmm\ RGS}\label{sect_rgs}

\begin{figure*}
\begin{center}
\includegraphics[width=0.8\textwidth]{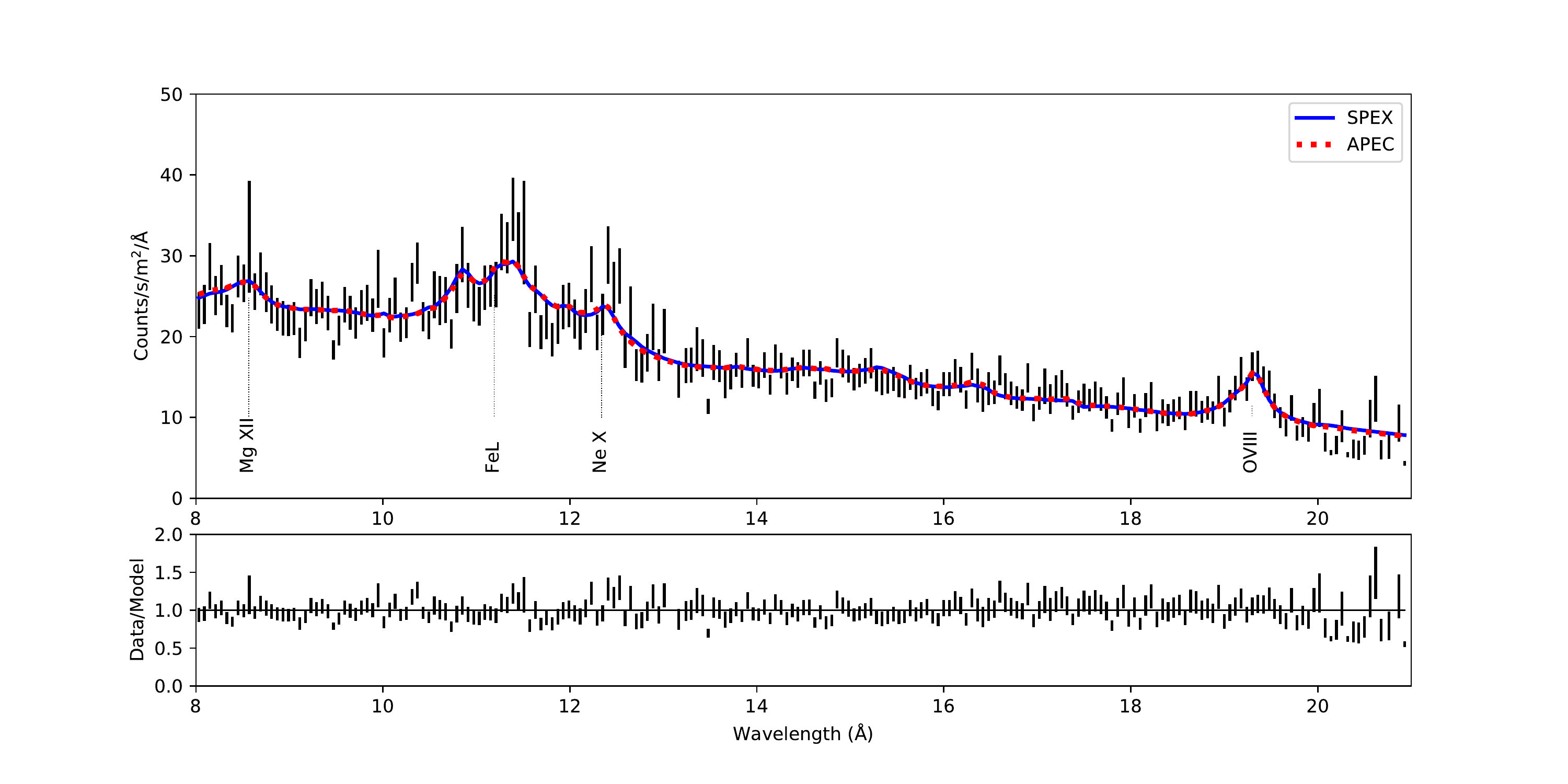} 
\end{center}
\caption{RGS data and best-fit models to the Perseus Cluster RGS spectra (SPEXACT in blue, solid and AtomDB in red, dotted lines). The ratio of data to model is shown with respect to the SPEXACT fit in the bottom panel. }\label{rgsfit}
\end{figure*}

The best-fit parameters obtained by performing a two-temperature fit to the \xmm\ RGS archival data are presented in Table \ref{tab_resrgs}. In Figure \ref{rgsfit}, the spectral models are compared to the stacked and binned data obtained from both detectors (RGS1 and RGS2) and from all observations used to constrain the fit. While the RGS extraction region is much narrower than the SXS FOV (0.8 compared to 3 arcmin), we consider it to offer the most reliable measurements of the typical properties of the Perseus Cluster core, since the spectral resolution is least degraded by the spatial extent of the source. 

Since the use of AtomDB within the SPEX spectral fitting package was only implemented very recently, following the launch of \hitomi, this offers us one of the first glimpses into directly comparing the two atomic line emission data bases in the soft X-ray band using grating spectra. While Figure \ref{rgsfit} shows that SPEXACT and AtomDB produce nearly indistinguishable model curves, the best-fit {\it parameters} are somewhat different, with AtomDB giving higher temperatures, lower normalisations, and a slightly improved C-stat compared to SPEXACT. In addition, while the Fe measurements agree for both models, we notice important differences in the best-fit abundance ratios, with AtomDB giving O/Fe and Ne/Fe values that are roughly 30\% higher than SPEXACT, while the Mg/Fe values are consistent between the two codes. Within the statistical error bars and the systematic uncertainty introduced by differences between the atomic codes, all measured ratios are consistent with the Solar composition, in line with the results for heavier elements from Si to Ni revealed by \hitomi. We have furthermore checked that the O/Fe, Ne/Fe, and Mg/Fe ratios obtained by fitting the second order RGS spectra are in agreement with the first order spectra presented here. 

\begin{table}
\caption{Best-fit spectral model parameters for the Perseus Cluster core, obtained from archival \xmm\ RGS spectroscopy. The temperatures (denoted $kT$) are expressed in keV, while the emission measures, $Y=\int n_e n_H {\rm d}V$, with $n_e$ and $n_H$ the electron and proton particle densities, are given in units of $10^{72}$ m$^{-3}$.}\label{tab_resrgs}
\begin{center}
\begin{tabular}{ccc}
\hline
 & APEC & SPEX \\
 \hline
O/Fe &  $1.31\pm0.19$ & $0.96\pm0.20$  \\
Ne/Fe  & $1.20\pm0.23$ & $0.78\pm0.21$  \\
Mg/Fe  & $0.84\pm0.18$ & $0.97\pm0.25$  \\
Ni/Fe  & $1.32\pm0.40$ & $1.28\pm0.40$  \\
\hline
Fe & $0.46\pm0.05$  & $0.42\pm0.06 $ \\
\hline
$kT_1$ &  $2.72\pm0.06$& $2.42\pm0.08$  \\
$kT_2$ & $0.60\pm0.02$ & $0.54\pm0.04$ \\
$Y_1$ & $6.08\pm0.53$ & $6.64\pm0.78$  \\
$Y_2$ & $0.11\pm0.01$ & $0.13\pm0.02$  \\
\hline
C-stat/d.o.f. & 4308.77/3616 & 4329.15/3616 \\
\hline
\end{tabular}
\end{center}
\end{table}

We note several discrepancies between the RGS and \hitomi\ SXS best-fit parameters. Firstly, as discussed in detail in the Tpaper, the best-fit thermal structure for the two data sets is different, with the RGS spectra preferring much lower temperatures than the SXS. This due to a combination of factors, including the different energy bands that the two detectors are sensitive to, different effective spatial extraction regions, and cross-calibration uncertainties. Secondly, due to the limited bandwidth of the RGS and thus the lack of line-free continuum, the absolute abundance of Fe is strongly degenerate with respect to the normalisation of the CIE component(s); this would explain the lower Fe abundance measured with the RGS compared to the SXS. Abundance ratios however, such as O/Fe, Ne/Fe, Mg/Fe, should not suffer from this degeneracy and are more reliable than the absolute RGS abundance measurements. We refer the reader to \citet{deplaa2017} for a comprehensive discussion of the systematic uncertainties associated with measuring these metal abundance ratios from RGS spectra.

\subsection{Summary of the chemical composition in the Perseus Cluster core measured from high-resolution spectroscopy}\label{sect_res_sum}

\begin{figure*}
\begin{center}
\includegraphics[width=1.6\columnwidth]{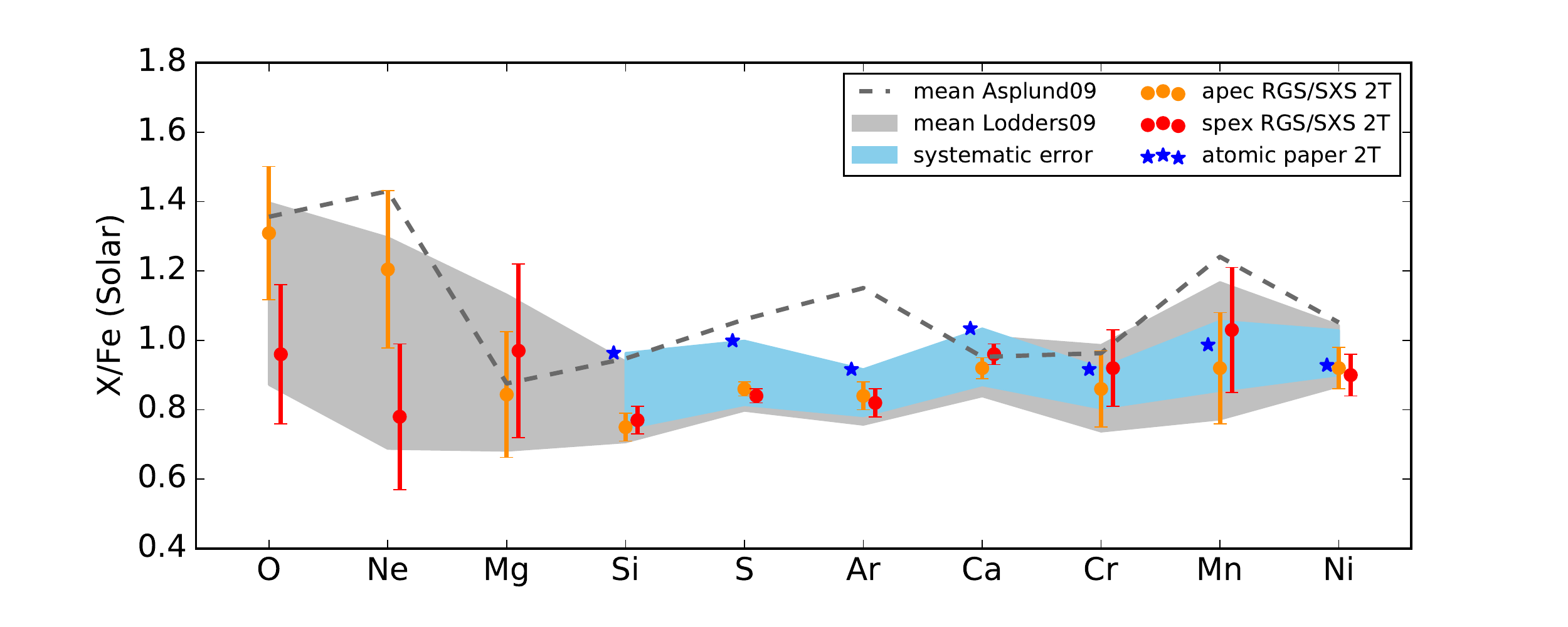} 
\end{center}
\caption{Summary of the metal abundance ratios with respect to Fe obtained using only the highest spectral resolution data sets available (O, Ne, Mg/Fe from \xmm\ RGS, Si through Ni to Fe ratios from \hitomi\ SXS). The light blue strip represents the systematic uncertainty on the \hitomi\ measurements from the `entire core' region due to the differences between AtomDB and SPEXACT and the effective area calibration; the inferred confidence range, including these systematic errors as well as the statistical uncertainties, is shown as a grey strip. The measured values are given with respect to the proto-solar units of \citet{lodders2009}; for comparison, the dashed grey line shows what the best-fit mean abundance ratios (i.e. the mid-points of the grey strip) would be if expressed instead in the Solar units of \citet{asplund2009}. }\label{fig_abundsum}
\end{figure*}

To obtain one unified set of abundances that can be compared to predictions from supernova yield calculations, we have combined the results presented above (O, Ne, Mg from RGS, and Si through Ni from the SXS `entire core' region). 
As shown in the Tpaper, the multi-phase thermal structure in the \hitomi\ spectrum is consistent with the expected effects of projecting the radial temperature gradient known to exist in the cluster core along the line of sight; hence, the `2 CIE' models provide a better description of the data (both in terms of the fit quality and as a better approximation of the physical characteristics of the cluster) and we will focus exclusively on these results for the remainder of the manuscript. Figure \ref{fig_abundsum} shows the best-fit metal abundance ratios using the `standard' ARF, assuming a `2 CIE' AtomDB and SPEXACT model. 

As mentioned above, the assumed effective area calibration introduces a systematic uncertainty in the results. In addition to tests with the `ground' and `crab' ARF shown in Table \ref{hitomiztab}, the Atomic paper presents an alternative method wherein residual calibration errors on the effective area are accounted for using correction functions whose parameters are estimated within SPEX (see Appendix 1.2 of that paper), while on the other hand assuming the continuum and line temperatures to be the same (unlike the approach used in the Tpaper and adopted here). For comparison, the results of the two-temperature model given in the Atomic paper are also shown in Figure \ref{fig_abundsum}.

\begin{table}
\caption{Summary of the abundance ratio constraints for the Perseus Cluster core, obtained from \xmm\ RGS and the \hitomi\ SXS `entire core' region. Uncertainties represent combined systematic and statistical errors at the 68\% confidence level.}\label{tab_ressum}
\begin{center}
\begin{tabular}{ccc}
\hline
& O/Fe &  $1.13\pm0.26$   \\
RGS & Ne/Fe  & $0.99\pm0.30$   \\
& Mg/Fe  & $0.91\pm0.22$  \\
\hline
& Si/Fe & $0.82\pm0.11$ \\
& S/Fe & $0.89\pm0.09$\\
& Ar/Fe & $0.84\pm0.08$ \\
SXS & Ca/Fe & $0.93\pm0.09$ \\
& Cr/Fe & $0.86\pm0.12$ \\
& Mn/Fe & $0.97\pm0.20$\\
 & Ni/Fe & $0.96\pm0.09$ \\
\hline
\end{tabular}
\end{center}
\end{table}

We hence estimate the total uncertainty budget for each of the measured abundance ratios as follows. For O, Ne, Mg, we averaged the abundances obtained from AtomDB and SPEXACT, and added in quadrature to the statistical uncertainty a systematic uncertainty term corresponding to the differences due to the assumed atomic data base. Typical uncertainties obtained in this way are of order 20--30\%, comparable to or larger than the expected systematic errors related to specifics of the RGS data analysis (estimated at around 20\% by \citealt{deplaa2017}). For the abundances constrained from the SXS, we calculate the average among seven different values for each element (a combination of two spectral codes with ARF `standard', `ground', or 'crab', plus the Atomic paper data points, assuming a two-temperature model in all cases). We then add in quadrature to the statistical uncertainty a systematic uncertainty term corresponding to the difference between the minimum and maximum values among the seven measurements. Figure \ref{fig_abundsum} and Table \ref{tab_ressum} summarise the confidence interval for each metal abundance ratio obtained in this way. 

The excellent quality of the \hitomi\ SXS data allows the abundance ratios with respect to Fe to be measured with a remarkable precision of less than 10\% for Ar/Fe, Ca/Fe, and Ni/Fe, less than 15\% for Si/Fe, S/Fe, and Cr/Fe, and 20\% for Mn/Fe, when all systematic uncertainties are taken into account. It is noteworthy that the abundances of some elements are measured more precisely than the Solar composition itself. Figure \ref{fig_abundsum} illustrates the differences between the \citet{lodders2009} and \citet{asplund2009} Solar reference units. In particular for Ar/Fe, the SXS constraints are much tighter than the systematic uncertainty associated with the assumed Solar abundance table. Hence, although X-ray spectra do not allow us to measure the abundances of various isotopes independently, our results can serve as an important benchmark for testing theoretical supernova yield calculations. This is discussed in detail in Section \ref{sect_yields}.

\subsection{Results from CCD spectroscopy}

\begin{figure*}
\begin{center}
\includegraphics[width=0.8\textwidth]{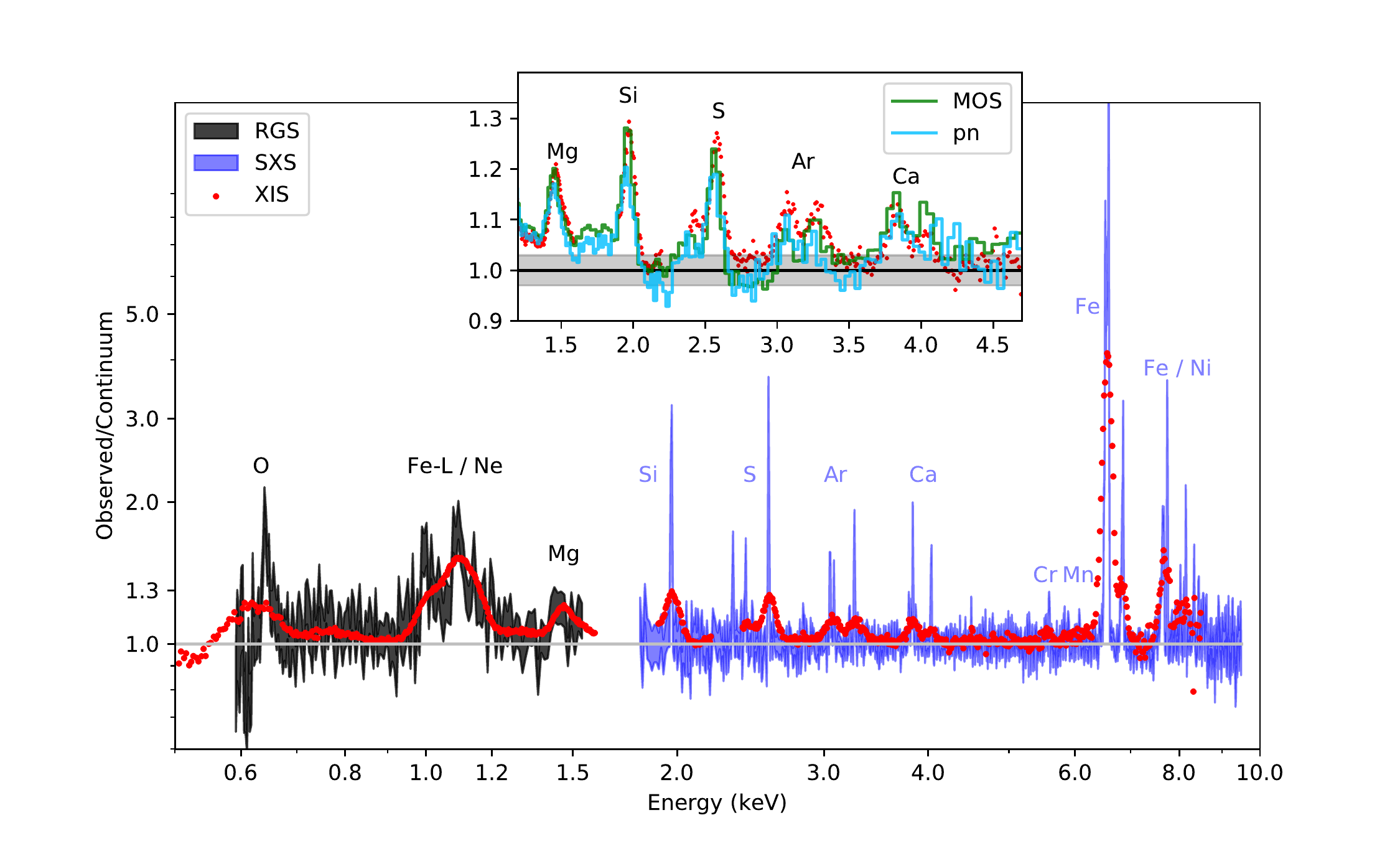} 
\end{center}
\caption{RGS(black), SXS(blue), and XIS(red,dotted) spectra of the core of the Perseus Cluster, after having subtracted the best-fit continuum models. The inset also shows a comparison between MOS (green), pn (light blue), and XIS (red,dotted) in the energy band containing the Mg, Si, S, Ar, Ca emission lines. The grey band corresponds to fluctuations at the $\pm$3\% level compared to the continuum.}\label{sxsrgsresid}
\end{figure*}

Prior to the launch of \hitomi, the workhorses of chemical abundance studies in galaxy clusters were primarily \xmm\ and \suzaku.
In this section, we compare the metal abundance constraints obtained from high-resolution spectroscopy to results from CCD detectors onboard these two satellites.
For both \suzaku\ and \xmm\ EPIC data, we reanalysed all available observations of the Perseus Cluster core using the latest atomic line emission data base updates, as well as an extraction region matched to the \hitomi\ FOV (roughly the `entire core' region). This allows us to perform the closest possible direct comparison between the results, varying only the type of detector while using the same atomic model and sky region wherever possible. We note that, following `standard' analysis methods used in previous chemical enrichment studies, the effects of resonant scattering were not modelled for \xmm\ and \suzaku\ data (but are expected to be small compared to other systematic uncertainties; see Table 1 of the Atomic paper and Figure 3 of ZNpaper).
Figure \ref{sxsrgsresid} shows the count residuals obtained by subtracting the line-free continuum from the SXS, RGS, EPIC, and XIS data, illustrating the X-ray signal used to derive our results and comparing the data quality in each case. 

\begin{table*}
\caption{Best-fit parameters for a region corresponding to the Perseus Cluster core, obtained from the archival CCD data sets considered in this work. Spectral normalisations $Y$ for \suzaku\ are given in standard XSPEC units normalised to an extraction area of $400\pi$ arcmin$^2$, while those for \xmm\ are quoted using the SPEX convention ($10^{72}$ m$^{-3}$). The most recent releases of AtomDB (v3.0.9) and SPEXACT (v3.03.00) were used in all cases.}\label{tab_res}
\begin{center}
\begin{tabular}{|c|ccc|cc|cc|}
\hline
 &  \suzaku & & & EPIC/MOS & & EPIC/pn & \\
\hline
  & APEC & SPEX & SPEX & SPEX & SPEX & SPEX & SPEX \\
  &  &   & local & & local & & local \\
\hline
O/Fe & $1.63\pm0.01$ & $1.32\pm0.01$ & $1.25\pm0.05$ & -- & $0.52\pm0.02$ & -- & $0.43\pm0.04$ \\
Ne/Fe  & $1.47\pm0.01$ & $0.95\pm0.01$ & -- & -- & -- & -- & -- \\
Mg/Fe   & $0.95\pm0.01$ & $0.91\pm0.01$ & $0.95\pm0.02$ & $0.99\pm0.02$ & $1.06\pm0.03$ & $0.27\pm0.06$ & $0.47\pm0.05$ \\
Si/Fe  & $1.01\pm0.01$ & $1.02\pm0.01$ & $1.02\pm0.01$ & $0.98\pm0.01$ & $0.99\pm0.01$ & $0.83\pm0.02$ & $1.04\pm0.02$\\
S/Fe  & $1.02\pm0.01$ & $0.95\pm0.01$ & $0.97\pm0.02$ & $0.72\pm0.02$ & $1.01\pm0.03$ & $0.79\pm0.02$ & $1.14\pm0.06$ \\
Ar/Fe  & $1.23\pm0.02$ & $1.24\pm0.02$ & $1.13\pm0.03$ & $0.73\pm0.04$ & $1.20\pm0.06$ & $0.84\pm0.07$ & $0.86\pm0.11$ \\
Ca/Fe & $0.83\pm0.02$ & $0.90\pm0.02$ & $1.05\pm0.04$ & $1.47\pm0.05$ & $1.30\pm0.06$ & $1.39\pm0.08$ & $1.21\pm0.12$\\
Cr/Fe  & $1.15\pm0.08$ & $1.10\pm0.08$ & $1.53\pm0.12$ & $1.20\pm0.18$ & $0.76\pm0.19$ & $1.49\pm0.28$ & $<0.76$ ($2\sigma$)\\
Mn/Fe & $1.45\pm0.11$ & $2.52\pm0.15$ & $1.89\pm0.25$ & $2.89\pm0.47$ & $1.67\pm0.44$ & $3.81\pm0.50$ & $0.99\pm0.69$ \\
Ni/Fe  & $0.87\pm0.02$ & $0.76\pm0.02$ & $1.39\pm0.06$ & $1.27\pm0.03$ & $1.00\pm0.14$ & $0.90\pm0.04$ & $1.13\pm0.16$\\
\hline
Fe & $0.821\pm0.001$ & $0.842\pm0.001$ & -- & $0.803\pm0.003$& -- & $0.659\pm0.003$ & -- \\
\hline
$kT_1$ & $3.327\pm0.002$ & $2.708\pm0.003$ & -- & $2.24\pm0.01$ & -- & $1.72\pm0.01$& -- \\
$kT_2$ & $6.32\pm0.02$ &  $5.159\pm0.007$ & -- & $4.99\pm0.03$ & -- & $4.67\pm0.02$ & -- \\
$Y_1$  & $10.973\pm0.004$ & $7.50\pm0.12$ & -- &  $5.25\pm0.11$ & -- & $3.43\pm0.07$& --  \\
$Y_2$  & $2.60\pm0.11$ & $6.677\pm0.004$ & -- & $9.02\pm0.07$ & -- & $9.65\pm0.07$ & -- \\
\hline
C-stat/d.o.f. & 125740.7/116172 & 126009.6/116172 & -- & 4518.77/1185 & -- & 2217.99/479 & -- \\
\hline
\end{tabular}
\end{center}
\end{table*}

The best-fit parameters obtained using the spectral analysis and modelling methods presented in Sections \ref{epic_analysis} (\xmm\ EPIC) and \ref{xis_analysis} (\suzaku\ XIS) are summarized in Table \ref{tab_res}. Because Ne lines are blended within the Fe-L complex, no `local fits' are reported for the Ne/Fe values. Moreover, as noted in Section \ref{epic_analysis},  for the \xmm\ EPIC analysis, Ne/Fe was tied to the RGS measurement, and O abundances were coupled between MOS and pn in the `global' fit. We thus do not report `global' Ne/Fe and O/Fe for \xmm.

Table \ref{tab_res} shows a large spread in abundance ratios obtained from various fitting methods (`local' vs. `global' fitting) and various CCD detectors (MOS, pn, and XIS). 
We have also performed an additional test in which we re-fitted the MOS and pn spectra in `local' energy bands while fixing the temperatures and ratio between the emission measures of the two components to the best-fit values obtained from Suzaku, and leaving only an overall normalization as free parameter. We find that differences between various CCD results persist, with a roughly similar spread as before, and therefore the temperature structure cannot be the leading cause of these discrepancies. Instead, the inset of Figure \ref{sxsrgsresid} shows the differences between the MOS, pn, and XIS signals around the Mg, Si, S, Ar, and Ca lines, illustrating why fluctuations as small as $\pm$3--5\% in the continuum level can lead to different line fluxes being measured from various fitting methods (`local' vs. `global') and from different instruments.
This highlights the importance of using a high-spectral resolution detector like the SXS for performing precise measurements of the chemical enrichment history of the ICM; a more quantitative assessment of the systematic uncertainties associated with CCD spectroscopy, based on the measurements given in Table \ref{tab_res}, is discussed in Section \ref{disc_sys}. 

\section{Discussion}

\subsection{Implications of a Solar chemical enrichment pattern in the core of the Perseus Cluster}\label{sect_solar}

\begin{figure*}
\begin{center}
\includegraphics[width=1.6\columnwidth]{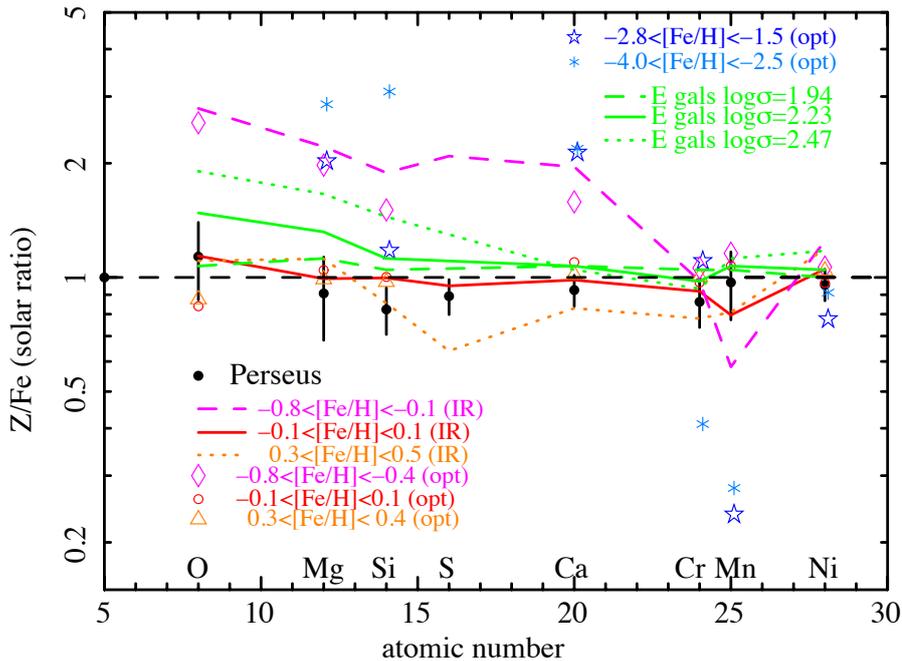} 
\end{center}
\caption{The metal abundance ratios of O/Fe, Mg/Fe, Si/Fe, S/Fe, Ca/Fe, Cr/Fe, Mn/Fe and Ni/Fe and the confidence levels of the Perseus Cluster (closed circles with error bars; plotted values correspond to those given in Table \ref{tab_ressum}). 
For comparison, we show the typical abundance patterns in early-type galaxies of various stellar velocity dispersion (log$\sigma$) obtained from stacked SDSS spectra \citep[results from][]{conroy2014}, as well as the average of Milky Way stars with various [Fe/H] derived from infrared spectra \citep{hawkins2016} and optical spectra \citep{bensby2014,battistini2015,reggiani2017,jacobson2015}.
All values were converted to the Solar abundance reference units of \citet{lodders2009} used throughout this work.}\label{stars}
\end{figure*}

The metal abundance ratios summarised in Section \ref{sect_res_sum} are remarkably close to the composition of the Sun. In fact, a `model' in which the chemical composition of the core of the Perseus Cluster is exactly Solar (using the reference values of \citealt{lodders2009} adopted throughout this work) provides an excellent match to the X-ray data, with a $\chi^2$ value of 10.7 for 10 degrees of freedom (d.o.f.) and no free parameters. This is different from the metal abundance ratios typically observed in the most massive galaxies in the Universe, such as central BCGs. As shown by \citet{conroy2014}, the relative abundances of $\alpha$ elements, such as O, Mg, and Si, with respect to Fe, increase as a function of the stellar velocity dispersion, $\sigma_*$, and stellar mass of a galaxy. Massive early type galaxies with velocity dispersion $\sigma_*\gtrsim200$~km~s$^{-1}$ have typically high $\alpha$/Fe ratios (up to twice the Solar value), which are inconsistent with the composition observed in the ICM of the Perseus Cluster core, as Figure \ref{stars} demonstrates. These ratios decrease with decreasing galaxy mass, reaching near-solar values for less massive galaxies, with $\sigma_*$ approaching 100~km~s$^{-1}$. 

The standard interpretation of [$\alpha$/Fe] is that this ratio is sensitive to the length of the star formation period, with higher values corresponding to shorter timescales. In giant ellipticals, which have very short star-formation timescales of $<1$~Gyr, and perhaps as short as 0.2 Gyr \citep{conroy2014}, there is not enough time for a significant fraction of SN~Ia to pollute the interstellar medium (ISM) before the end of the starburst period. Although SN~Ia continue to produce metals later on, these elements are no longer incorporated into stars once star formation ends, leading to a high [$\alpha$/Fe] ratio seen in stellar spectra. Galaxies with younger stellar populations and longer periods of star formation, on the other hand, show a chemical composition closer to the Solar value and to the abundance pattern observed in the ICM of the Perseus Cluster core.

This trend in [$\alpha$/Fe] as a function of the timescale/level of chemical enrichment can be seen even more clearly among individual stars in the Milky Way, whose abundances can be studied more directly and with fewer assumptions than by using integrated spectra of more distant galaxies. 
In Figure \ref{stars}, we also provide a comparison between the abundance ratios measured in the Perseus Cluster using X-ray spectroscopy and the average composition of stars in the Milky Way derived from a large spectroscopic infrared survey \citep{hawkins2016}, and high-resolution optical surveys \citep{bensby2014, battistini2015, reggiani2017, jacobson2015}. The chemical enrichment pattern observed in the Perseus Cluster core is remarkably similar to the abundance ratios of solar-metallicity stars in our Galaxy, while lower-metallicity stars (which also formed much earlier than the Sun) show significantly enhanced [$\alpha$/Fe] ratios. The Ca/Fe measurements in particular can be considered the most robust, since results from different stellar observations are consistent, and the [Ca/Fe] vs. [Fe/H] trend provides a good match to the Galactic chemical evolution model \citep{reggiani2017}. The plateau of [Ca/Fe] in metal-poor stars at twice the Solar ratio indicates that SNcc contribute significantly towards the production of Ca; as the relative contribution from SN~Ia and consequently [Fe/H] increase, the $\alpha$-element to Fe ratio becomes closer to the Solar value. The \hitomi\ SXS provides a remarkably precise measurement of the Ca abundance in the ICM of the Perseus Cluster, and the resulting Ca/Fe ratio is in excellent agreement with that of Solar-metallicity stars.  

Unlike [$\alpha$/Fe], the ratios of Fe-group elements such as Cr/Fe, Mn/Fe, and Ni/Fe are close to the Solar composition, showing no dependence on the stellar metallicity (for Milky Way stars), or on the galaxy stellar mass (for the sample of galaxies considered by \citealt{conroy2014}). Early hints that the Mn/Fe ratio of Galactic stars was decreasing with decreasing metallicity, possibly indicating a different origin for Mn than the other Fe-group metals, were likely attributed to the effects of non local-thermodynamical-equilibrium \citep{battistini2015}. 
The good agreement between the Galactic stars, the early-type galaxy sample of \citealt{conroy2014}, and the Perseus Cluster ICM further indicates that the production of Cr, Mn, and Ni traces very closely that of Fe, independently of the exact details of the chemical enrichment history. This is expected if these Fe-group elements are all predominantly produced through the same nucleosynthesis channel (namely, SN~Ia explosions with similar progenitor properties).

As summarised in Section~\ref{intro}, recent results (using either moderate-resolution CCD spectroscopy or high-resolution spectroscopy based on a few elemental ratios only) point towards the emerging picture of an ICM chemical composition that is consistent with Solar and does not vary significantly with radius, with [Fe/H], or from cluster to cluster. Our present results, compiling the most accurate abundance measurements that can be constrained in the Perseus Cluster to date, are the latest addition in this line of recent evidence, as they support both a Solar composition of the ICM (as shown above), and a lack of spatial gradient in the chemical enrichment pattern of all observed elements (Section \ref{sect_radial}). Unfortunately, our measurements are limited to the central core (within $\sim$100~kpc), containing a small fraction of the total ICM mass, and representing an area within which the mass of metals trapped in stars is comparable to that in the gas component. Precise measurements of the metal abundance ratios over a larger volume and in a larger sample of systems are needed to further improve our understanding of the chemical enrichment history of the ICM as a whole. 

Nevertheless, it is worth noting that retrieving Solar abundance ratios in the vicinity of Perseus Cluster's BCG is surprising, at least at first glance. The stellar population of the BCG itself, as argued above, is expected to have super-solar [$\alpha$/Fe] ratios which could contribute to the enrichment of the ICM through stellar winds; on the other hand, SNcc are very rare in early-type galaxies \citep{graham2012}, such that the metal output by these old stellar populations is dominated by SN~Ia \citep{mannucci2008}, reducing the [$\alpha$/Fe] in the ICM. It would be remarkable if these two enrichment sources compensated each other to such a precision that the composition of the ICM remains Solar, even within the small error bars afforded by the \hitomi\ SXS. However, one could also argue that, when systems have a similar initial stellar mass function and are old enough, such that most supernovae (both early core-collapse and longer time scale SNIa) have already exploded, the abundance pattern might be expected to be similar, even if the instantaneous rate of release of metals into the surrounding gas has been different from system to system over time. 

\subsection{Constraints on the contribution from different supernova yield models to the enrichment of the ICM}\label{sect_yields}

\begin{table*}
\caption{Summary of the theoretical supernova yield calculations considered in this work.}\label{tab_yieldsum}
\begin{center}
\begin{tabular}{cclccc}
\hline
Model  & SN  & Description & Dimen- & Metallicity  & Reference \\
name & type &  & sions & ($Z_\odot$) &  \\
\hline
W7 & SNIa & near-$M_{Ch}$ (single degenerate), deflagration & 1D & 1.0 & \citet{iwamoto1999} \\
WDD1,2,3 & SNIa & near-$M_{Ch}$, delayed detonation & 1D & 1.0 & \citet{iwamoto1999} \\ 
W7new & SNIa & like W7, updated electron capture and nuclear reaction rates & 1D & 1.0& \citet{nomoto2018} \textsuperscript{\ref{xx}} \\
WDD2new & SNIa & like WDD2, updated electron capture and nuclear reaction rates & 1D & 1.0 & \citet{nomoto2018} \textsuperscript{\ref{xx}} \\ 
300-Z-c3-1 & SNIa & near-$M_{Ch}$, delayed detonation & 2D &  0, 0.5, 1.0 & \citet{leung2017} \\
500-1-c3-1 & SNIa & like 300-Z-c3-1,  higher progenitor central density & 2D &  1.0 & \citet{leung2017} \\
N100 & SNIa & near-$M_{Ch}$, delayed detonation & 3D & 1.0 & \citet{seitenzahl2013} \\
\hline
`1.1-0.9' & SNIa & double degenerate, both sub-$M_{Ch}$ WD explode & 3D & 1.0 & \cite{pakmor2012} \\
DDDDDD & SNIa & only one sub-$M_{Ch}$ WD in a double-WD binary explodes & 3D & 1.0 & \cite{shen2018} \\
\hline
Nomoto & SNcc & thermally driven explosion & 1D & 0.2, 1.0 & \cite{nomoto2006} \\
SukN20 & SNcc & including neutrino transport, calibrated so that a  & 1D & 1.0 & \cite{sukhbold2016} \\
 & &  \citet{nomoto1988} progenitor explodes like SN1987A & &  & \\
SukW18 & SNcc & including neutrino transport, calibrated so that a & 1D & 1.0 & \cite{sukhbold2016} \\
 & & \citet{utrobin2015} progenitor explodes like SN1987A & &  & \\
\hline
\end{tabular}
\end{center}
\end{table*}

\begin{table*}
\caption{Results of fitting a combination of SNcc and SNIa yield models to the metal abundance ratios observed in the core of the Perseus Cluster. When $f_{\rm Ia}$ is the only free parameter in the fit, the $\chi^2$ value is minimised using only the O/Fe, Ne/Fe, and Mg/Fe measurements; when both $f_{\rm Ia}$ and $f_{\rm ch}$ are allowed to vary, O/Fe, Ne/Fe, Mg/Fe, Cr/Fe, Mn/Fe, and Ni/Fe are used to constrain the fit parameters. The model obtained in this way is then extrapolated to the other abundance measurements not used in the fit procedure (denoted $\chi^2$/all). For a small subset of model combinations, we also show the results obtained if all 10 abundance ratios are used to constrain the fit; in this case, the two $\chi^2$ columns are merged.}\label{tab_fIa}
\begin{center}
\begin{tabular}{lccccc}
\hline
\hline
SNIa model & SNcc model & $f_{\rm Ia}$ & $f_{\rm ch}$ & $\chi^2$/d.o.f. & $\chi^2$/all \\
\hline
W7 & Nomoto $Z_\odot$  SP & $0.20\pm0.04$ & -- & 0.61/2 & 180.65 \\
WDD1 &  Nomoto $Z_\odot$  SP & $0.22\pm0.04$ & -- & 0.63/2 & 103.47  \\
WDD2 &  Nomoto $Z_\odot$  SP & $0.19\pm0.03$ & -- & 0.66/2 & 39.83 \\
WDD3 &  Nomoto $Z_\odot$  SP & $0.17\pm0.03$ & -- & 0.67/2 & 31.91 \\
\hline
WDD2new &  Nomoto $Z_\odot$  SP & $0.20\pm0.04$ & -- & 0.64/2 & 52.99 \\
W7new &  Nomoto $Z_\odot$  SP & $0.19\pm0.03$ & -- & 0.64/2 & 30.38 \\
\hline
W7new &  Nomoto $Z_\odot$  TH & $0.25\pm0.04$ & -- & 0.71/2 & 37.11 \\
W7new &  Nomoto 0.2$Z_\odot$  SP & $0.22\pm0.03$ & -- & 0.09/2 & 35.09 \\
\hline
W7new &  Nomoto $Z_\odot$  SP & $0.15\pm0.01$ & -- & \multicolumn{2}{c}{25.91/9 } \\
\hline
\hline
300-0-c3-1 & Nomoto $Z_\odot$  SP & $0.18\pm0.03$ & -- & 0.68/2 & 14.33 \\
300-0.01-c3-1 & Nomoto $Z_\odot$  SP & $0.19\pm0.03$ & -- & 0.67/2 & 19.01 \\
300-0.02-c3-1 & Nomoto $Z_\odot$  SP & $0.19\pm0.03$ & -- & 0.67/2 & 30.90 \\
300-0-c3-1 & Nomoto 0.2$Z_\odot$  SP & $0.21\pm0.03$ & -- & 0.09/2 & 54.82 \\
\hline
300-0-c3-1 & Nomoto $Z_\odot$  SP & $0.21\pm0.02$ & -- & \multicolumn{2}{c}{11.78/9 } \\
\hline
\hline
N100 + DDDDDD & Nomoto $Z_\odot$  SP & $0.25\pm0.06$ & $0.36\pm0.14$ & 7.97/4 & 23.96 \\
500-1-c3-1 + '1.1-0.9' & Nomoto $Z_\odot$ SP & $0.27\pm0.06$ & $0.13\pm0.09$ & 7.32/4 & 33.59 \\
N100 + '1.1-0.9' & Nomoto $Z_\odot$  SP & $0.25\pm0.06$ & $0.30\pm0.14$ & 3.23/4 & 28.16 \\
\hline
N100 + DDDDDD & Sukhbold W18 & $0.33\pm0.06$ & $0.12\pm0.11$ & 6.49/4 & 40.24 \\
N100 + DDDDDD & Sukhbold N20 & $0.34\pm0.06$ & $0.11\pm0.11$ & 5.04/4 & 17.21 \\
N100 + '1.1-0.9' & Sukhbold N20 & $0.36\pm0.07$ & $0.11\pm0.11$ & 3.59/4 & 24.74 \\
\hline
N100 + DDDDDD & Sukhbold N20 & $0.38\pm0.06$ & $0.09\pm0.09$ & \multicolumn{2}{c}{15.73/8 }  \\
\hline
\hline
\end{tabular}
\end{center}
\end{table*}

Since we find such a good agreement between the chemical enrichment pattern in the Solar neighbourhood, typical Milky Way stars, and the ICM in the core of the Perseus Cluster, it is important to test how well current supernova yield calculations are able to reproduce this abundance pattern that seems to pervade the Universe. To this end, we construct a model consisting of a linear combination of theoretical SN~Ia and SNcc yields from the literature, and compare it to the observations. The various theoretical supernova yield calculations considered here are listed in Table \ref{tab_yieldsum}, and described below.

As possible model yields for SN~Ia, we use: 
\begin{enumerate}
\item older calculations of ``classical'' SN~Ia yields, focusing on the most widely used WDD1,2,3 and W7 models of \citet{iwamoto1999};
\item a recent update of the WDD2 and W7 models using revised electron capture and nuclear reaction rates \citep{nomoto2018} \footnote{http://supernova.astron.s.u-tokyo.ac.jp/\~{}nomoto/yields \label{xx}};
\item two-dimensional delayed detonation near-Chandrasekhar mass ($M_{Ch}$) explosion models from the 300-Z-c3-1 series of \citet{leung2017}, in order to study the effect of progenitor metallicity Z on the metal yields; 
\item the latest three-dimensional calculations for a delayed-detonation of a near-$M_{Ch}$ progenitor \citep[model N100 from][]{seitenzahl2013};
\item as an alternative to the N100 model, we also test the 500-1-c3-1 model from \citet{leung2017}, which assumes a higher central density of a single-degenerate progenitor such that neutron-rich species are generated more efficiently;
\item a double degenerate scenario consisting of a violent merger between two sub-$M_{Ch}$ white dwarfs, described by model `1.1-0.9' of \cite{pakmor2012};
\item a ``dynamically-driven double-degenerate double-detonation'' scenario (`DDDDDD'), in which only one of the white dwarfs involved in the collision is detonated (``mass = 1.0 M$_\odot$, C/O = 50/50, metallicity = 0.02, 12+16 = 1.0'' model from \citealt{shen2018});
\end{enumerate}
While a variety of other SN~Ia nucleosynthesis models is available in recent literature, the ones listed above match the properties of typical SN~Ia explosions (e.g., the mass of $^{56}$Ni and hence the maximum brightness) most closely. 

In terms of the core-collapse supernova yield calculations, we consider: 
\begin{enumerate}
\item the ``classical'' mass-dependent values of \citet{nomoto2006,nomoto2013} for various progenitor metallicities, averaged over an initial mass function (IMF) assumed to have a power-law shape. The most common assumption is that of a Salpeter IMF \citep{salpeter1955} with slope -2.35 (hereafter `SP'), but we also test a shallower, top-heavy IMF with a slope of -1.35 (hereafter `TH'). 
\item more recent calculations by \citet{sukhbold2016} which include a one-dimensional neutrino transport model for the explosion and are calibrated to reproduce the observed energy for SN 1987A (specifically, models W18 and N20 from that work).
\end{enumerate}

 
For a combination of a single SN~Ia and a single SNcc model, the fit has one free parameter representing the relative number of SN~Ia over the total number of SNe responsible for the enrichment (which we denote as $f_{\rm Ia}$). In addition, ZNpaper focused on the relative abundances of Fe-group elements determined from the SXS spectra to show that a mixture of sub- and near-$M_{Ch}$ SN~Ia progenitors is preferred. Therefore, we also consider linear combinations of two different types of SN~Ia progenitors together with a given SNcc yield model. This introduces a second possible free parameter, the fraction of near-$M_{Ch}$ SN~Ia explosions with respect to the total number of SN~Ia responsible for the enrichment (which we denote as $f_{\rm ch}$). The subsections that follow describe the constraints on these parameters, and the extent to which the models are able to reproduce the measured chemical enrichment pattern in the Perseus Cluster core determined in Section \ref{sect_res_sum}.


\subsubsection{``Classical'' supernova yield calculations}\label{yields_classical}

\begin{figure*}
\begin{center}
\includegraphics[width=1.5\columnwidth]{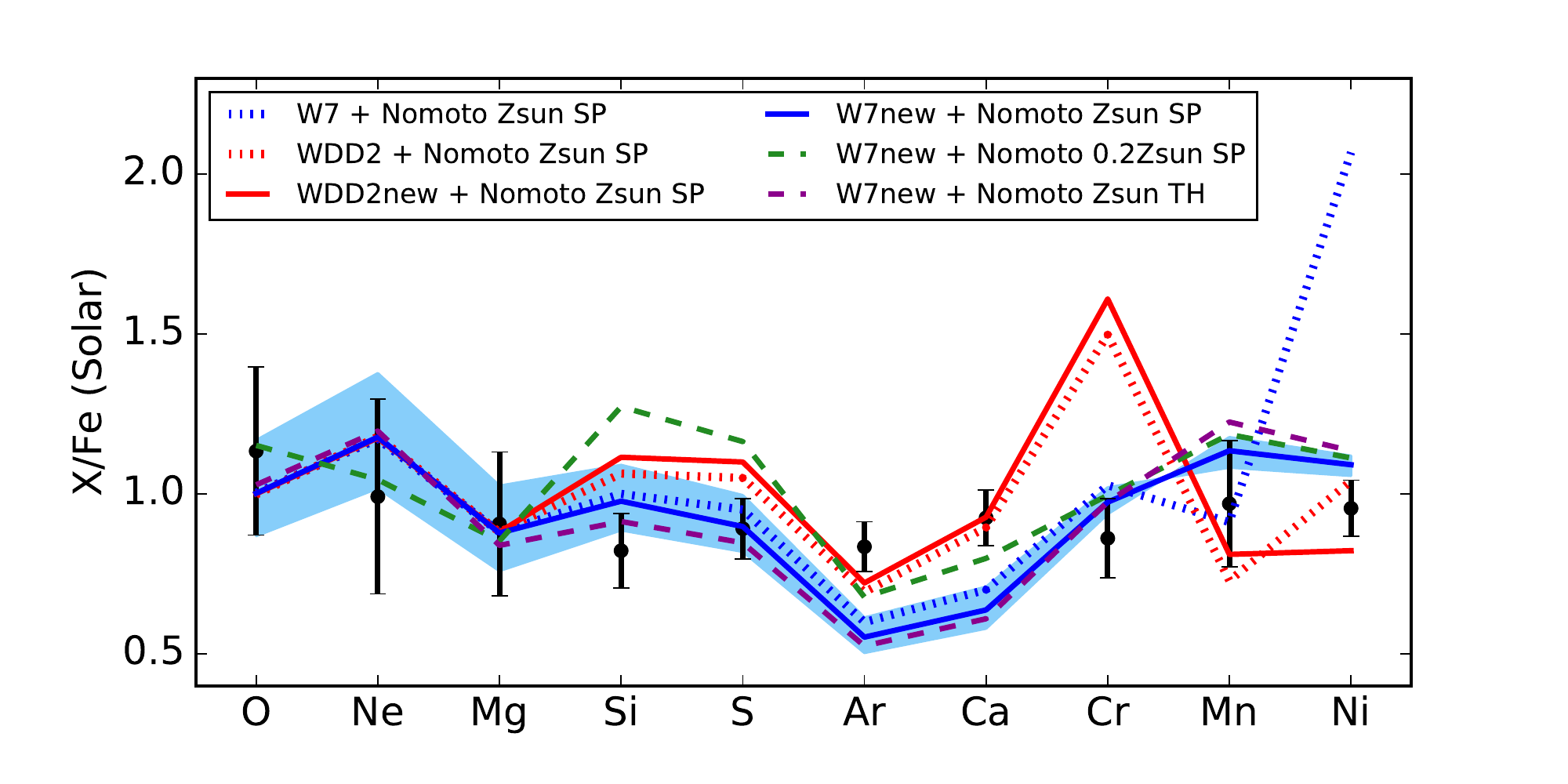} 
\includegraphics[width=1.5\columnwidth]{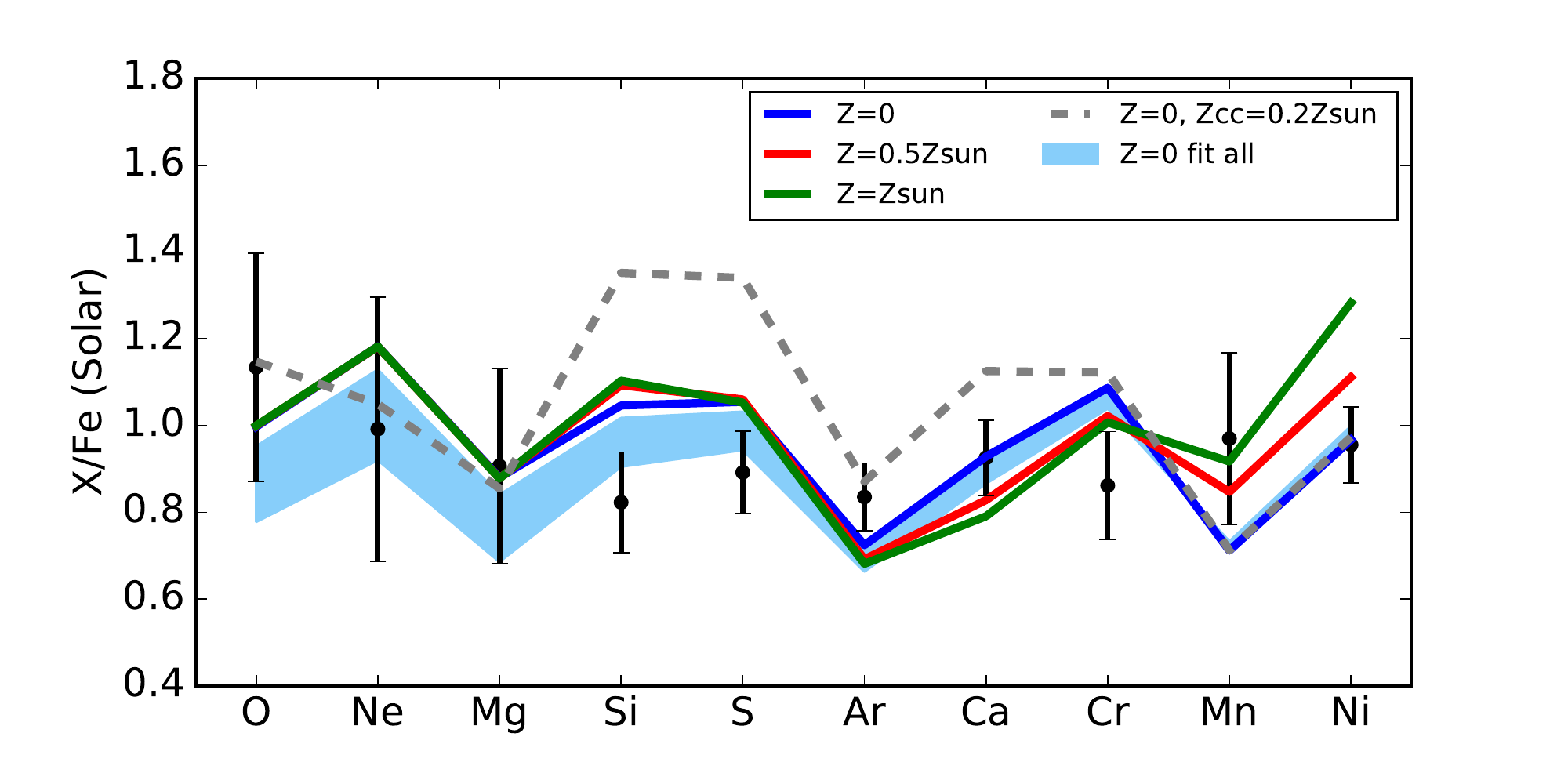}
\end{center}
\caption{{\it Top:} Models obtained using various combinations of `classical' SN~Ia and SNcc yields. The WDD2 and W7 SN~Ia yields are taken from \citet{iwamoto1999}; their updated calculations with revised electron capture and nuclear reaction rates by \citet{nomoto2018} are the labeled `new'. For SNcc, we use the calculations of \citet{nomoto2006,nomoto2013}, with Solar (`Zsun') or 0.2 Solar metallicity (`0.2Zsun'), and assuming a Salpeter (`SP') or top-heavy (`TH') IMF, as indicated by the labels.  {\it Bottom:} Solid lines assume the `Nomoto Zsun SP' SNcc yields, together with various SN~Ia progenitor metallicities based on the 300-Z-c3-1 series of \citet{leung2017}. The grey dashed line assumes a lower metallicity of the SNcc progenitor.  
In both panels, the black data points with error bars show the confidence range of the SXS/RGS measurements (Table \ref{tab_ressum}), and the blue band illustrates the effects of varying $f_{\rm Ia}$ within its allowed $1\sigma$ confidence interval. In the top panel, this $1\sigma$ confidence interval is determined fitting only O/Fe,Ne/Fe, and Mg/Fe, while in the bottom panel, we present the confidence interval obtained by fitting all data points.}\label{fig_fIa_classic}
\end{figure*}

First, we consider models consisting of a linear combination of the ``classical'' and widely used supernova yields from \citet{iwamoto1999} for SN~Ia and \citet{nomoto2006,nomoto2013} for SNcc, with $f_{\rm Ia}$ as the only free parameter. Since O, Ne, and Mg are mainly forged by SNcc, their relative abundances compared to Fe, which is a predominantly SN~Ia product, are most sensitive to $f_{\rm Ia}$. However, due to the different sensitivities of the RGS and SXS, the relative error bars on the O/Fe, Ne/Fe, and Mg/Fe ratios are significantly larger than the uncertainties in determining Si/Fe through Ni/Fe. This could lead to biases wherein the best-fit model misses the light $\alpha$-element to Fe measurements by a large margin in order to compensate for uncertainties/imperfections in the model yields. Hence, for the cases where $f_{\rm Ia}$ is the only free parameter, we have chosen to estimate its best-fit value using only O/Fe, Ne/Fe, and Mg/Fe. The constraints obtained in this way are summarised in Table \ref{tab_fIa}, showing both the $\chi^2$ value that was minimised in the fit procedure (in this case, for two degrees of freedom given by three data points and one free parameter), and the $\chi^2$ value obtained by extrapolating the resulting model to all 10 abundance measurements constrained by the SXS/RGS data; the top panel of Figure \ref{fig_fIa_classic} over-plots the model predictions and the observed chemical enrichment pattern in the core of the Perseus Cluster.
  
Within this set of models, the best-fit values obtained for $f_{\rm Ia}$ are in the range $21\pm4$\%, independent of the choice of exact SN~Ia and SNcc progenitor properties. Figure \ref{fig_fIa_classic} shows that significantly more precise measurements of the O, Ne, and Mg abundances are needed to distinguish between various assumptions employed for the SNcc yields. It is notable that, while slightly increasing the {\it relative number} of core-collapse supernovae that contribute to the enrichment, a top-heavy IMF has a barely noticeable effect on the resulting {\it abundance patterns}, which will be difficult to constrain even with a marked improvement in data quality. The initial metallicity of the SNcc progenitors on the other hand has a larger impact on the predicted O/Ne/Mg ratios than the assumed IMF; future observations can be used to constrain this parameter but, at present, the differences between models are still smaller than the error bars associated with the current measurements. Throughout the remainder of this manuscript, we will then focus on the `natural' choice of Solar progenitor metallicity and Salpeter IMF. Note that a top-heavy IMF is disfavoured by other considerations, such as the observed metal-mass-to-light ratios in galaxy clusters \citep{matsushita2013}.

While the observed O/Fe, Ne/Fe, and Mg/Fe ratios are reproduced well by any choice of models in Figure \ref{fig_fIa_classic}, several noteworthy discrepancies can be seen for other elements.

In terms of the Fe-group metals, it has long been proposed that the nickel-to-iron ratio in the ICM can be used as a discriminator between SN Ia explosion models, with high Ni/Fe pointing towards a slow deflagration (W7) model, whereas delayed detonation scenarios would result in a lower value of this ratio \citep{dupke2000}. Figure \ref{fig_fIa_classic} indeed shows the dramatic difference in predicted Ni/Fe for the ``classical'' SN~Ia yields of \citet{iwamoto1999}. 
It is important to note, however, that updating the electron capture and nuclear reaction rates results in a substantially different Ni production. While deflagration models still result in a larger Ni/Fe compared to delayed detonation models, this difference is reduced significantly compared to the older calculations of \citet{iwamoto1999}. The largest difference between the `W7new' and `WDD2new' model \citep{nomoto2018} is now in the Cr/Fe ratio, which is much higher for the delayed detonation than for the deflagration explosion. 

Among the models tested in this subsection, the \citet{nomoto2006} SNcc calculations with Salpeter IMF and Solar initial metallicity of the progenitor, combined with the updated W7 SN~Ia deflagration model of \citet{nomoto2018}, provide the best description of the observed Perseus abundance pattern. However, comparing this model to the 10 metal abundance ratios constrained by the SXS/RGS data, we still obtain a high (extrapolated) $\chi^2$ value of 30.38. For this model combination, we also provide a fit obtained by using all measured abundance ratios to minimise the $\chi^2$. The fit is still not acceptable, with $\chi^2=25.91$ for 9 d.o.f., and the value of $f_{\rm Ia}$ is biased low compared to the fit using only the light $\alpha$-element to Fe ratios. This model combination over-predicts all Fe-group to Fe abundances compared to the observed data.

In addition, the abundances of intermediate $\alpha$-elements are also not reproduced well by the models considered so far. In particular, Si/Ar and Si/Ca are always over predicted, such that Ar and Ca are underproduced when Si is well-fit, and Si is overproduced when Ar and Ca are well fit. Similar fit residuals for these intermediate elements have been reported from previous CCD measurements of the cluster sample presented by \citet{mernier2016ii}, as well as earlier work by \citet{deplaa2007}. Several solutions have been discussed to explain the deficit of calcium produced in theoretical supernova models compared to values measured in the ICM using \xmm\ data. \citet{deplaa2007} suggest that using the yields from an empirically modified delayed-detonation model calibrated on the Tycho type Ia supernova remnant improves the agreement between the predicted and observed Ca/Fe. Alternatively, a possible subclass of SNIa called ``calcium-rich gap transients'' may also contribute to the enrichment of the ICM \citep{mulchaey2014}. However, delayed detonation models tend to significantly over produce Cr (see above), and calcium-rich gap transients only have a very minor contribution to Ar production \citep{mernier2016ii}. Hence, these previous explanations are not likely to address the problem that the observed Si/Ar ratio is lower than the model predictions for all cases tested in Figure \ref{fig_fIa_classic}. 

These ``classical'' yield calculations, though still widely used as baseline models in the literature until recently, were proposed almost two decades ago. Since then, there have been recent improvements not only in the electron capture and nuclear reaction rates (see above), but also in modelling 2D and 3D explosions (introducing possible asymmetries in their geometry) as well as in predicting the SN~Ia yield dependencies as a function of various properties of the progenitor and the explosion mechanism. Moreover, for the best-fit values of $f_{\rm Ia}$ presented here, the contribution from SNcc always represents more than 50\% of the total Si, S, Ar, Ca budget predicted by the models. Updates to SNcc nucleosynthesis calculations may therefore be equally important for improving the agreement with the observed abundance pattern in the ICM. These start-of-the-art models are considered in the next subsections.

\subsubsection{Dependence on the SN~Ia progenitor metallicity}

As shown in the previous section, ``classical'' SN~Ia yield models do not reproduce well the observed Fe-group to Fe abundance ratios, both in the case of delayed-detonation and deflagration models, even when the latest electron capture and nuclear reaction rates are applied. Here, we investigate whether relaxing the assumption of a Solar SN~Ia progenitor metallicity can lead to an improvement in the fit results. To this end, we begin by considering combinations of the \citet{nomoto2006} SNcc calculations with Salpeter IMF and Solar initial metallicity, together with the delayed detonation single degenerate explosion models from the 300-Z-c3-1 series of \citet{leung2017} (where Z stands for the assumed SN~Ia progenitor metallicity). Following the argument presented above, we rely on the light $\alpha$-element to Fe ratios to constrain $f_{\rm Ia}$ as the only free parameter in the fit. The results are given in Table \ref{tab_fIa} and illustrated in the bottom panel of Figure \ref{fig_fIa_classic}. While the goodness-of-fit for the Solar progenitor metallicity SN~Ia model of the 300-Z-c3-1 series is similar to that obtained using the classical `W7new' yields above (with a $\chi^2$ of 30.90 compared to 30.38 for `W7new'), calculations with a lower initial metallicity of the SN~Ia result in lower Mn/Fe and Ni/Fe (and a slight increase in Cr/Fe), providing overall a better description of the data. Note that a similar trend of the Mn/Fe and Ni/Fe as a function of initial metallicity of the SN~Ia progenitor is also found in the 3D calculations of \citet{seitenzahl2013}.
Figure \ref{fig_fIa_classic} also shows that the Cr/Fe ratio is significantly reduced between the 1D (e.g. `WDD2new') and 2D (300-1-c3-1) delayed detonation models, even for the same initial progenitor metallicity.
The best fit is obtained by mixing the \citet{nomoto2006} SNcc calculations with Salpeter IMF and Solar initial metallicity and the zero metallicity model (300-0-c3-1) of \citet{leung2017}. In fact, this combination provides a statistically acceptable description of the measurements; including all data points to constrain the fit, we obtain $\chi^2=11.78$ for 9 d.o.f. 

Nevertheless, it seems somewhat unphysical to assume that the SN~Ia and SNcc progenitor metallicity would be so different. Models where both the initial metallicity of the SNcc and the SN~Ia are sub-Solar, on the other hand, provide significantly worse fits: assuming SNcc progenitors with 0.2$Z_\odot$, together with 0 metallicity SN~Ia, gives a clearly unacceptable $\chi^2=54.82$ (mainly due to a very high Si and S compared to O, Ne, and Mg). 
Given that several other lines of evidence had previously suggested an early enrichment scenario in which many of the SN~Ia that enrich the ICM exploded already during the proto-cluster phase \citep[e.g.][]{werner2013,mantz2017}, it is tantalising to consider the possibility that these SN~Ia progenitors may have indeed had a highly sub-Solar metal content. However, the Fe-group to Fe abundance ratios are sensitive to several other factors in addition to the progenitor metallicity -- in particular, various scenarios for sub- and near-$M_{Ch}$ SN~Ia explosions can affect the predicted yields significantly. In the following sections, we focus on searching for alternative model combinations that can explain the observed abundance pattern without assuming a large discrepancy between the SN~Ia and SNcc progenitor metallicities.

\subsubsection{Combinations of near- and sub-$M_{Ch}$ SN~Ia progenitors}\label{yields_mix}

\begin{figure*}
\begin{center}
\includegraphics[width=1.5\columnwidth]{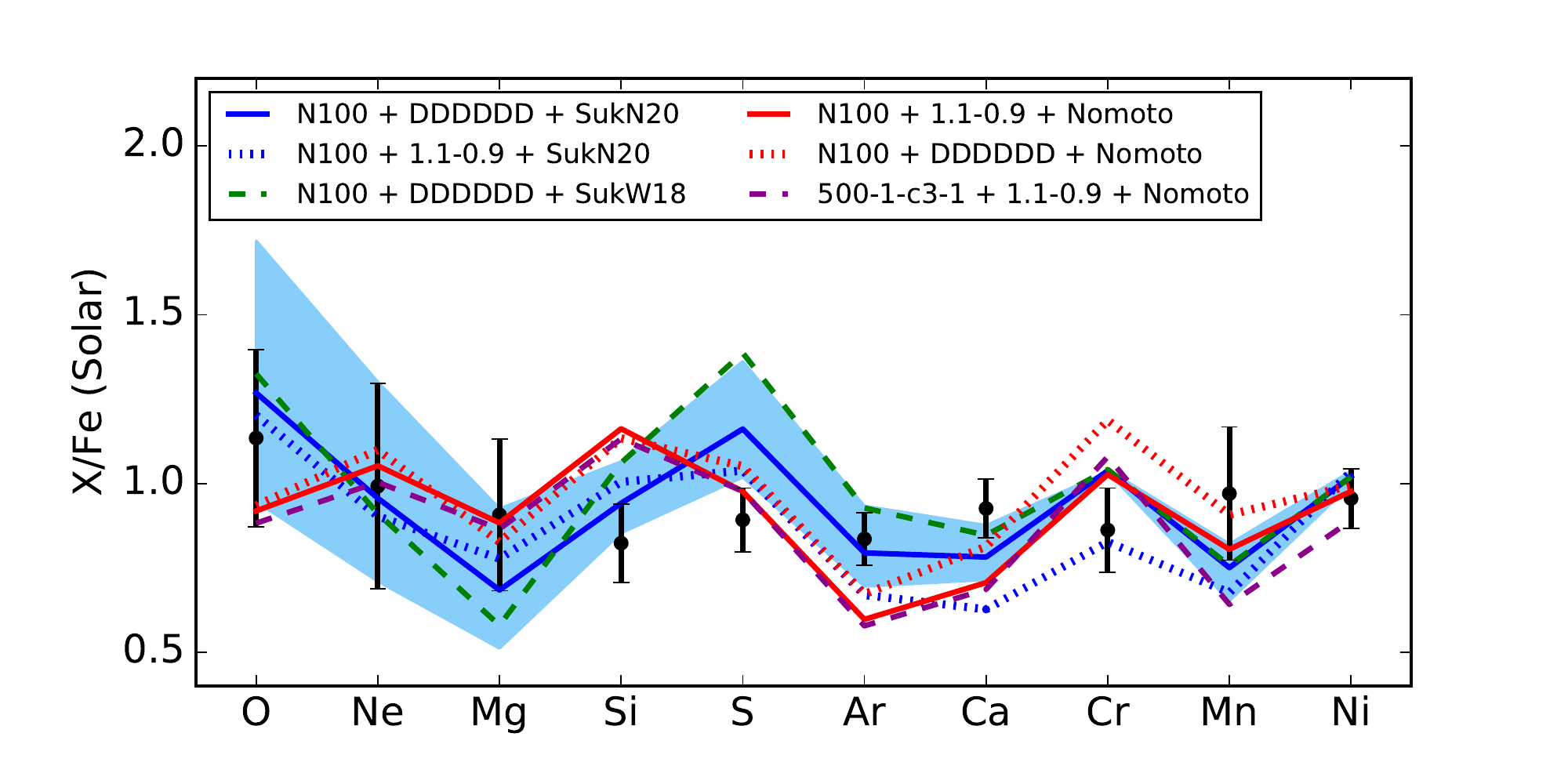} 
\includegraphics[width=\columnwidth]{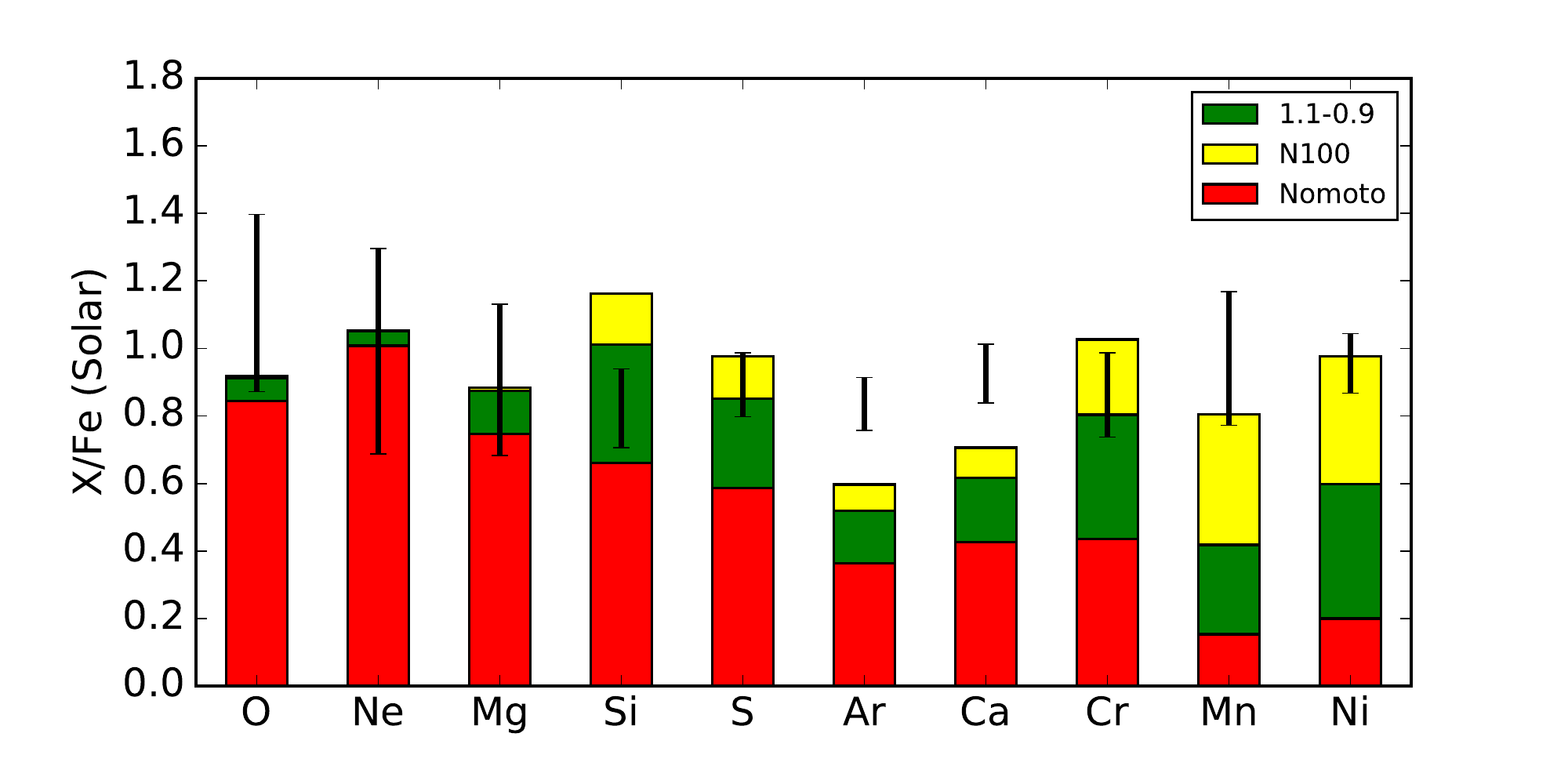} 
\includegraphics[width=\columnwidth]{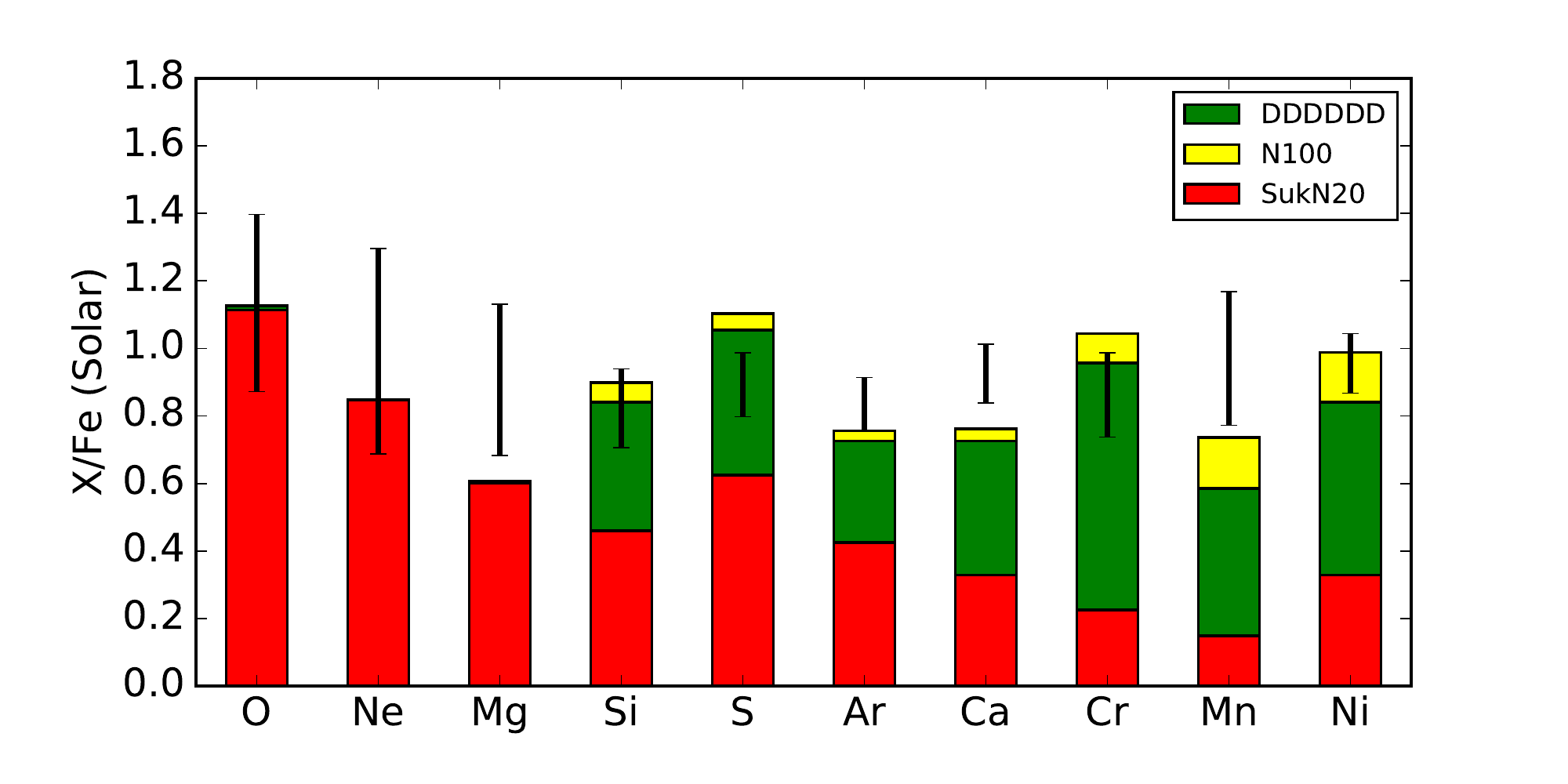} 
\end{center}
\caption{Results of fitting a combination of near- and sub-$M_{Ch}$ yield models for SN~Ia, together with various assumptions for SNcc. All progenitor metallicities are assumed to be Solar. For near-$M_{Ch}$ SN~Ia, we use the models N100 from \citet{seitenzahl2013} and 500-1-c3-1 from \citet{leung2017}; for sub-$M_{Ch}$ SN~Ia, we test the calculations of \citet{pakmor2012} (model `1.1-0.9') and \citet{shen2018} (labeled `DDDDDD'). For SNcc, models `SukN20' and `SukW18' are from \citet{sukhbold2016}, and `Nomoto' from \citet{nomoto2006}. SNcc yields are integrated assuming a Salpeter IMF. 
In the top panel, we compare the results of fitting only the O/Fe, Ne/Fe, Mg/Fe, Cr/Fe, Mn/Fe, and Ni/Fe ratios with various model combinations (the models were extrapolated to predict the abundances of intermediate elements from Si to Ca). The blue band illustrates the effects of varying $f_{\rm Ia}$ and $f_{\rm ch}$ within their allowed $1\sigma$ confidence interval.
In the bottom panels, bars of different colours show the contribution from various SN types to the enrichment model for two particular cases: on the left, the combination that gives the best description of the Fe-group abundances; on the right, the combination that gives the best description of all abundance ratios, including intermediate elements. For the latter case only (the bottom right plot), \textit{all} abundance measurements were used to constrain the best-fit model parameters.}\label{fig_fIa_bar}
\end{figure*}

As argued in ZNpaper, linear combinations of sub- and near-$M_{Ch}$ SN~Ia progenitors also provide a good match to the Fe-group to Fe abundance ratios observed in the Perseus Cluster. While the ZNpaper assumed a 1:1 ratio between these two explosion channels, here we attempt to constrain their relative contribution more quantitatively. To this end, we model the O/Fe, Ne/Fe, Mg/Fe, Cr/Fe, Mn/Fe, and Ni/Fe data points with $f_{\rm Ia}$ and $f_{\rm ch}$ as free parameters. The ratios of light elements with respect to Fe should determine the best-fit $f_{\rm Ia}$, while the abundances of Fe-group elements should be relatively insensitive to the choice of SNcc model and should constrain $f_{\rm ch}$. The abundances of intermediate elements (Si to Ca), for which the dependance on $f_{\rm Ia}$ and $f_{\rm ch}$ is much more complex, are not used in the fit, unless stated otherwise. The results for various sub-$M_{Ch}$ and near-$M_{Ch}$ models are summarised in Table \ref{tab_fIa} and illustrated in Figure \ref{fig_fIa_bar}.

Assuming the metal production from SNcc to follow the calculations of \citet{nomoto2006} with Salpeter IMF and Solar initial metallicity, the best-fit to the O/Fe, Ne/Fe, Mg/Fe, Cr/Fe, Mn/Fe, and Ni/Fe data points is given by a combination of the N100 delayed detonation model for the near-$M_{Ch}$ progenitors, and the `1.1-0.9' violent merger model for the sub-$M_{Ch}$ explosion channel. The fit describes the six considered data points well (with a $\chi^2$ of 3.23 for 4 d.o.f and two free parameters) and suggests $f_{\rm Ia}=0.25\pm0.06$ (in agreement with results presented in the previous sections) and $f_{\rm ch}=0.30\pm0.14$, i.e. about one third of SN~Ia progenitors are near-$M_{Ch}$.
However, this model still suffers from the same problems in reproducing the relative abundances of intermediate $\alpha$-elements as the `classical' models in the previous section. In fact, because the total mass of the two white dwarfs involved in the collision described by the `1.1-0.9' model is relatively high, a lot of unbound material undergoes incomplete burning, resulting in a large amount of Si (and even the production of some lighter elements from O to Mg, which are normally not expected to be contributed by SN~Ia in an appreciable quantity; see bottom left panel of Figure \ref{fig_fIa_bar}). Hence, the Si/Ar ratio predicted by this model is even higher, and in worse tension with the observations, than the `classical' models discussed above. In the DDDDDD scenario, on the other hand, only one of the WDs involved in the collision explodes, leaving the other one intact. The mass of unbound ejecta is smaller, resulting in a smaller amount of light elements being produced. This leads to an overall better match to the observed Perseus Cluster metal abundance pattern when all elements (including Si-Ca) are considered. However, Si/Ar and Si/Ca are still higher than the observed values (though the tension is less than when using the N100+`1.1-0.9' model). In addition, the DDDDDD yields over predict Cr, such that the goodness of fit when considering only the O/Fe, Ne/Fe, Mg/Fe, Cr/Fe, Mn/Fe, and Ni/Fe data points is worse than in the N100+`1.1-0.9' case. 

Finally, we note that the best-fit value of $f_{\rm ch}$ is sensitive to the choice of near-$M_{Ch}$ SN~Ia model (while it seems to be robust to the choice of sub-$M_{Ch}$ model). Recent X-ray observations of the supernova remnant 3C 397 indicated Mn/Fe and Ni/Fe ratios much higher than those predicted by typical near-$M_{Ch}$ SN Ia models \citep{yamaguchi2015}, suggesting that near-$M_{Ch}$ white dwarfs might have a higher central density than previously thought \citep{dave2017,leung2017}. Using a higher density near-$M_{Ch}$ SN Ia model (`500-1-c3-1' from \citealt{leung2017}) as an alternative to the N100 model, in combination with the `1.1-0.9' sub-$M_{Ch}$ model and the \citet{nomoto2006} SNcc yields, we obtain a lower $f_{\rm ch}=0.13\pm0.09$. This combination of models however provides a worse fit than the other two combinations described above, both when considering only the O/Fe, Ne/Fe, Mg/Fe, Cr/Fe, Mn/Fe, and Ni/Fe data points, and when extrapolating the model to all Perseus measurements including Si,S,Ar,Ca. 

\subsubsection{The impact of different SNcc model calculations}\label{yields_sncc}

All combinations of SN~Ia yields discussed above show the same residuals in fitting the intermediate $\alpha$-elements from Si--Ca when the SNcc yields are assumed to follow the model of \citet{nomoto2006}. It is therefore possible that the differences between the observed data and the models are due to remaining uncertainties in SNcc nucleosynthesis calculations.

Recently, \citet{sukhbold2016} presented an alternative set of SNcc nucleosynthesis calculations taking into account neutrino transport. The models presented by \citet{sukhbold2016} were calibrated by adjusting the model parameters such that, when the explosion engine is applied to a progenitor known to roughly match that of SN 1987A, the resulting energy and nickel mass also agree with the observational constraints for that event. With the parameters determined in this way, the explosion model is then applied to a suite of Solar-metallicity progenitors with a wide range of masses from 9 to 120 $M_\odot$ evolved from the main sequence. This calibration leads to different results depending on the exact progenitor model assumed for SN 1987A; here we focus on two choices, namely the N20 and W18 progenitor models.

N20 is a historical model wherein a main-sequence star of 21.0 $M_\odot$ evolves into a red-supergiant, undergoes mass loss to become a 16.3 $M_\odot$ star, and then transforms into a blue-supergiant as a result of extra artificial dredge-up of He into the envelope \citep{saio1988}.  The star forms a 6.0 $M_\odot$ He core, so that the luminosity and radius are consistent with the observed progenitor of SN 1987A.  The evolution of the 6.0 $M_\odot$ He core is calculated separately through the early collapse of the Fe core \citep{nomoto1988}. The explosion model of the 16.3 $M_\odot$ star satisfies the observational constraints on the SN 1987A explosion \citep{shigeyama1990}. Model W18 on the other hand is based on an 18 $M_\odot$ star that is evolved intact from the main sequence; the assumed high rotation leads to mixing of He into the envelope, such that the star develops into a blue supergiant resembling the progenitor of SN 1987A; however, this evolution leads to a larger He core than suggested by the SN 1987A explosion \citep{utrobin2015}.  

We have assumed a Salpeter IMF to compute the average of the mass-dependent yields reported by \citet{sukhbold2016}, calibrated either to a W18 or N20 progenitor. The contribution from both the explosion and pre-SN winds is included. The results are shown in Table \ref{tab_fIa} and Figure \ref{fig_fIa_bar}. Both the W18 and N20 models produce significantly more O, and a higher O/Mg ratio, than the yields of \citet{nomoto2006}; as a result, in order to reproduce the O/Fe, Ne/Fe, and Mg/Fe measurements in the core of the Perseus Cluster, the contribution from SNcc assuming the \citet{sukhbold2016} yields must be reduced, i.e. the best-fit $f_{\rm Ia}\sim35\pm6$\% is significantly higher than when using the SNcc yields of \citet{nomoto2006}. The N20 and W18 models provide a better fit to the Si/Ar and Si/Ca ratios compared to the SNcc model of \citet{nomoto2006}, when mixed with a given combination of SN~Ia products. However, the W18 model overproduces S by a very large factor and is hence disfavoured by the data. 

Fitting {\it all} metal abundance ratios measured with the SXS/RGS with a combination of the \citet{sukhbold2016} N20 yields, averaged over a Salpeter IMF, the N100 yields for near-$M_{Ch}$ SN~Ia, and DDDDDD yields for sub-$M_{Ch}$ SN~Ia, gives an acceptable goodness of fit, with a $\chi^2$ of 15.73 for 8 d.o.f. and $f_{\rm Ia}=38\pm6$\%. This fit does not require a significant contribution from near-$M_{Ch}$ SN~Ia, with $f_{\rm ch}$ spanning allowed values between 0 and 20\%. We note however that the Mn/Fe ratio is not reproduced well, and that when intermediate elements from Si to Ca are excluded from the fit, the best $\chi^2$ is obtained by combining the \citet{nomoto2006} SNcc yields with the N100 and `1.1-0.9' models for SN~Ia (see previous subsection).

Therefore, no single combination of the models considered above clearly stands out as the best solution for explaining the metal abundance ratios observed in the core of the Perseus Cluster. Although mixing the \citet{nomoto2006} SNcc yields for a Solar progenitor metallicity with the \citet{leung2017} delayed detonation near-$M_{Ch}$ SN~Ia model for zero metal abundance of the progenitor gives a formally acceptable fit, the discrepancy required between the SNcc and SNIa progenitor compositions makes this unlikely to be a physically reasonable model. If we require the SNcc and SNIa progenitors to have the same metal abundance, the \citet{sukhbold2016} N20 yields for SNcc, together with the ``dynamically-driven double-degenerate double-detonation'' sub-$M_{Ch}$ SN~Ia model, are preferred in order to reproduce the relative abundances of intermediate $\alpha$-elements, while the \citet{nomoto2006} SNcc yields, together with the N100 and `1.1-0.9' models for near- and sub-$M_{Ch}$ SN~Ia, provide the best match to the Fe-group elements. In addition, the lowest $\chi^2$ value obtained assuming SNcc and SNIa progenitors to have the same initial metallicity (15.73 for 8 d.o.f.) is still much worse than the simple assumption that all ratios are equal to Solar ($\chi^2=10.7$ for 10 d.o.f. and no free parameters).

In reality, it is likely that the picture is even more complex: Type Iax supernovae \citep{jha2017}, as well as Ca-rich gap transients \citep{waldman2011}, may still contribute to the enrichment. Given the uncertainties associated with current SN yields, it is difficult to quantify the contribution from these additional processes at this time. The fact that linear combinations of sets of  SN~Ia and SNcc models do not reproduce the observed abundances well may also be partly due to a failure in one of the implicit assumptions underlying the linear combination model. This relatively simple approximation of the enrichment assumes that a few classes of supernovae dominate the chemical enrichment, while in reality it may have taken several generations of stars with many different configurations to form the currently observed (Universal) abundance pattern in the first few billion years after the Big Bang. 

\begin{figure*}
\begin{center}
\includegraphics[width=1.6\columnwidth]{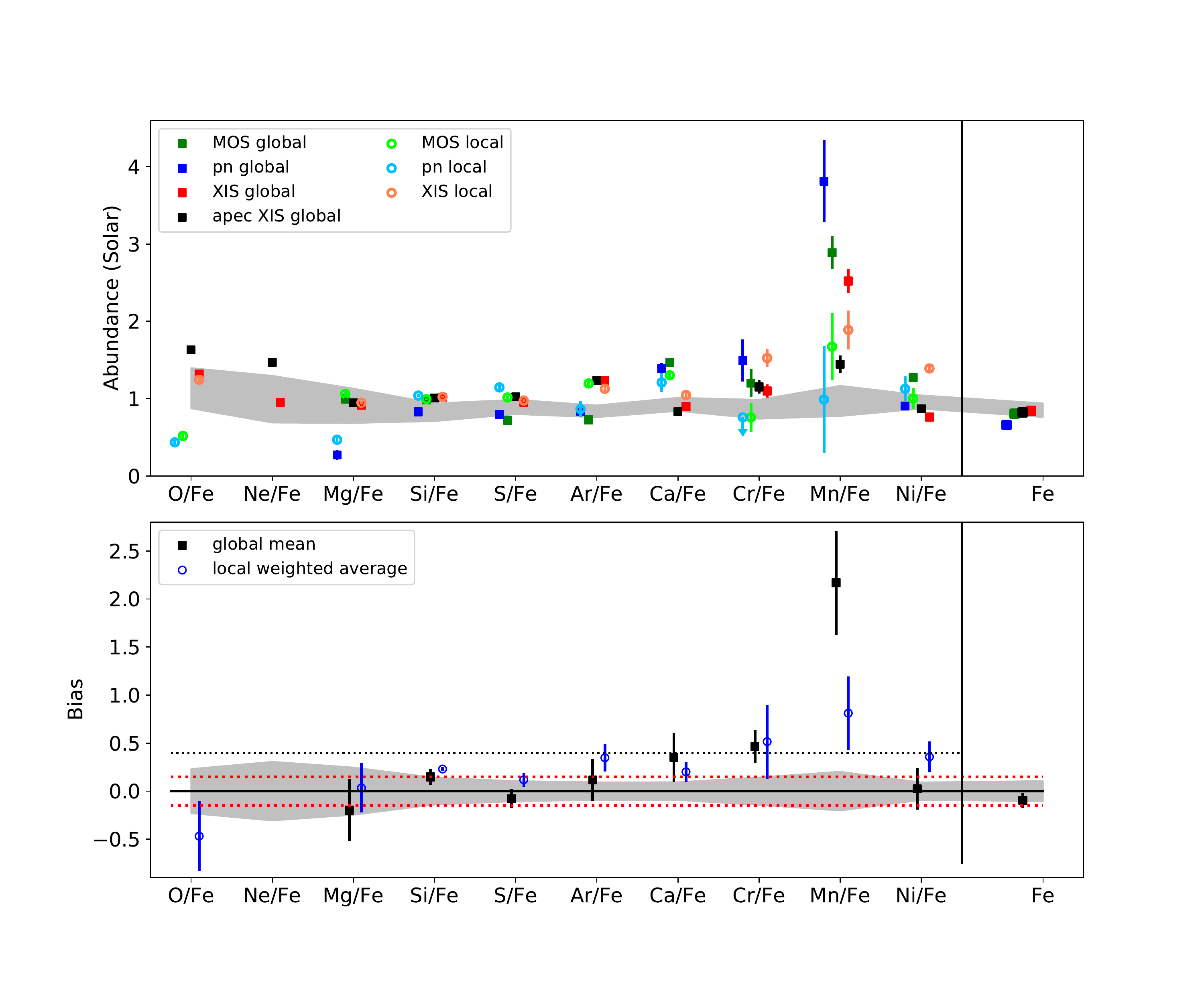} 
\end{center}
\caption{{\it Top:} abundance ratios in the core of the Perseus Cluster for a region corresponding to the \hitomi\ SXS FOV, measured with \suzaku\ XIS and \xmm\ EPIC for both `local' and `global' fits. For the `global' fits, absolute Fe/H measurements are also compared. Unless explicitly stated otherwise in the legend, the fit was performed using SPEXACT v3.03.00. The data points are based on the values shown in Table \ref{tab_res}. The confidence interval obtained using the higher spectral resolution SXS and RGS data is shown as a grey band.
{\it Bottom:} the average relative bias of CCD results (calculated as the difference between the CCD results and the SXS/RGS mean value, divided by the SXS/RGS mean value). Black squares are obtained by averaging the results of global fits to the CCD data (using only SPEXACT in order not to count the XIS results twice). Blue circles show the average of the local fits, weighted by their respective statistical errors. Error bars represent the standard deviation between XIS, MOS, and pn in each case. The grey band represents the confidence interval of the RGS/SXS measurements. The horizontal lines show no bias (solid black), $\pm$15\% bias (dotted red), and +40\% bias (dotted black).}\label{fig_bias}
\end{figure*}

\subsection{Implications for metal abundance studies using X-ray CCDs}\label{disc_sys}

The exquisite multi-observatory data available for the core of the Perseus Cluster allows us to perform a detailed comparison of the chemical abundance pattern measured using various detectors and fitting techniques. While ZNpaper discussed the differences between previously published studies using CCD data and the \hitomi\ results, here we have ensured that the archival observations are reprocessed such that the exact same spatial region and version of the atomic line emission data base is used, in order to provide the most direct comparison possible. Although any systematic uncertainties identified for this system may not be representative for spectra of other galaxy clusters with different redshifts and thermal structures, we hope that these results provide a useful gauge of the residual calibration errors associated with X-ray CCD spectroscopy. This is particularly relevant in this context, since biased abundance ratio measurements from previously utilised CCD data may result in incorrect constraints on SNIa and/or SNcc yield models (see Section \ref{sect_yields}).
In the top panel of Figure \ref{fig_bias}, we show the differences between the metal abundance ratios measured with \suzaku\ XIS, \xmm\ EPIC, and the higher spectral resolution \hitomi\ SXS and \xmm\ RGS. For the CCD measurements, the results of both `local' and `global' fits are compared, and described in more detail below.    

\begin{itemize}
\item The Fe/H abundances obtained with EPIC/MOS and \suzaku/XIS are in good agreement with the \hitomi\ SXS result (within about 5\%), while EPIC/pn and \xmm/RGS give significantly lower Fe/H values. This is a known issue with RGS analysis that should not have a large impact on the relative abundance ratios with respect to Fe (see Section \ref{sect_rgs}). For EPIC/pn, however, it is less clear how the best-fit metal abundance ratios are affected by this systematic difference in the absolute Fe abundance (which is about 20\% lower than the SXS measurement).
\item There is a large spread in the best-fit O/Fe values. The \suzaku\ XIS gives larger values of O/Fe than \xmm\ EPIC, regardless of whether a `global' or `local' fit method is used for \suzaku\ (for EPIC, only `local' fit results are reported, since a global fit was not performed independently for MOS and pn for this element). The RGS value falls between the XIS and EPIC measurements. Both the O emission line and the O absorption edge from the Galactic foreground, as well as the contamination layer calibration of \suzaku, can contribute to these systematic differences. Figure \ref{sxsrgsresid} shows very broad residuals around the O line in Perseus for the XIS, illustrating the difficulty of obtaining a reliable best-fit value. Additionally, AtomDB results predict a systematically larger O/Fe compared to SPEXACT.
\item The Ne abundance is uncertain since the strongest lines from this element are blended with the Fe-L complex, preventing `local' fitting methods from being applied. As for O, AtomDB fits give a larger Ne/Fe compared to SPEXACT. For a given assumption of the atomic data base, \suzaku\ measurements of Ne/Fe lie within the (relatively large) statistical confidence interval of the corresponding RGS measurements. For \xmm\ EPIC, the Ne/Fe ratio in the global fits was fixed to that obtained from RGS and is not reported.
\item The Mg/Fe abundance ratios measured with  \suzaku\ XIS and EPIC/MOS are in good agreement with \xmm\ RGS, while EPIC/pn underestimates the Mg/Fe value. 
\item The Si/Fe ratio obtained from nearly all CCD measurements is at the upper end of the confidence interval of the \hitomi\ SXS result, except the `global' EPIC/pn value that is lower and in better agreement with the SXS. In practice, given that the absolute Fe abundance measured with EPIC/pn is significantly lower than with the SXS (by about 20\%), this means that the {\it absolute} Si abundance is underestimated by the EPIC/pn global fit and overestimated by other methods, compared to the SXS. We caution however that, due to the effective area attenuation by the closed gate valve, the Si abundance is the most prone to systematic uncertainties among the elements detected by \hitomi. 
\item there are statistically significant differences between the `local' and `global' fit results for S using both EPIC detectors and for Ar using EPIC/MOS. The `global' best-fit values are typically lower than the `local' values; the SXS measurements for S/Fe and Ar/Fe lie between the EPIC/MOS `local' and `global' results, and are in agreement with the EPIC/pn `global' fits. 
The \suzaku\ XIS results for these elements are more stable with respect to differences between the `global' and `local' fits compared to the EPIC measurements. The S/Fe ratio measured by \suzaku\ is within the SXS confidence interval, while Ar/Fe is overpredicted by the XIS results.
\item The XIS measurements of Ca/Fe are in relatively good agreement with the \hitomi\ results, while all estimates of Ca/Fe from \xmm\ overestimate this ratio. This systematic uncertainty may at least in part be responsible for previous claims that Ca-rich gap transient contributions and/or alternative SNIa yield models were required to explain the metal enrichment pattern in the ICM (all papers discussing this effect were based on \xmm\ data; see \citealt{deplaa2007,mernier2016ii}). We note however that, even though the Ca/Fe ratio obtained with the SXS is lower than \xmm\, it remains difficult for many choices of ``classic'' supernova models to produce enough Ca to match the observations (see Section \ref{sect_yields}).
\item For both Cr/Fe and Ni/Fe, the best-fit `local' XIS value is in {\it worse} agreement with the SXS than the `global' fit. The `local' MOS value of Cr/Fe is in agreement with the \hitomi\ result, while Cr is not detected significantly in the EPIC/pn `global' fits.  Both the `local' and `global' Cr/Fe measurements obtained with XIS, as well as the `global' EPIC results, are higher than the \hitomi\ value.
\item Mn/Fe ratios are overestimated by all detectors and all fitting methods, with the possible exception of the `local' EPIC/pn result (the latter having large statistical uncertainties). The large discrepancy between atomic codes for the XIS-derived `global' Mn/Fe values is somewhat mitigated by using `local' fits, which significantly decrease the Mn/Fe value measured with SPEXACT, while the `local fitting' AtomDB result (Mn/Fe=$1.43\pm0.20$, not plotted in Figure \ref{fig_bias}) remains the same as the `global' value. In addition, at medium spectral resolution, the Mn lines may blend with Fe XXII -- Fe XXIV innershell transitions, known to be quite uncertain and to differ in strength between AtomDB and SPEXACT, contributing to the large uncertainty in the Mn measurement.
\item The Ni/Fe abundances obtained with both EPIC/pn and EPIC/MOS `local' fits are in good agreement with the \hitomi\ results, while the XIS `global' fits slightly underestimate, and the `local' fit overestimates the SXS measurement. Leaving both Ni and Fe to vary freely in the `local' XIS fit (i.e. using only the He-$\beta$ Fe line to measure the Fe abundance) on the other hand gives Ni/Fe=$0.80\pm0.06$, this time consistent with the \hitomi\ value. In any case we note that, employing the latest updates to the emission line data bases in AtomDB and SPEXACT, the high values of Ni/Fe reported by several previous studies of metal enrichment in the ICM \citep{dupke2000,mernier2016i} are no longer favoured.     
\item Except for O, Ne and Mn, abundances obtained with AtomDB and SPEXACT are in good agreement.
\end{itemize}


It seems therefore that all choices of detector (XIS, MOS or pn) and fitting methodology (`local' or `global') yield the `correct' result in the case of some elements, but fail to match the high-resolution spectroscopy results for other abundance ratios. Small residual `wiggles' in the effective area calibration present at various energies for different CCD instruments affect the results for that respective detector (see inset of Figure \ref{sxsrgsresid}). Allowing the overall normalisation in a narrow energy band to be free in the fit can only shift the continuum level up or down, but not correct for any `curvature' uncertainty in the effective area calibration (i.e. the continuum is too high on one side of the line but too low on the other), which is perhaps why `local' fits do not necessarily provide a better agreement with the micro-calorimeter data than `global' fits. 

To further quantify the average biases associated with CCD spectroscopy, we have also computed the mean and standard deviation of the results obtained with the different detectors considered here. In order not to count the XIS data set twice, we only use the SPEXACT results for this calculation (as noted above, for all abundances measured by the SXS, except Mn, SPEXACT and AtomDB agree very well; for Mn/Fe, the AtomDB measurement with XIS falls well within the spread in results obtained with SPEXACT for the various CCD detectors considered here). We compute the average values separately for the `global' and `local' fits. For the `global' fits, the statistical error bars are formally very small, and systematic errors clearly dominate; hence, we have computed a simple mean value and standard deviation for the MOS, pn, and XIS measurements. The `local' fits typically provide an acceptable description of the limited energy band in question (i.e. a reasonable goodness-of-fit); hence, the statistical errors in this case (denoted $\sigma_{\rm stat}$) can be considered as more reliable, and we have computed the average of the MOS, pn, and XIS measurements weighted by $1/\sigma_{\rm stat}^2$. The upper limit on Cr/Fe from the pn local fit is not used in the calculation. The results are shown in the bottom panel of Figure \ref{fig_bias}, where we plot the average relative bias (difference between the mean CCD and RGS/SXS measurement, divided by the RGS/SXS result) and standard deviation of the abundance ratios measured with lower-resolution spectra.

Within the scatter among various CCD detectors, the mean O/Fe and Mg/Fe ratios agree well with the current RGS constraints. Taking the \textit{average} value between \xmm\ and \suzaku\ results, the `global' Fe/H, Si/Fe, S/Fe, Ar/Fe, and Ni/Fe ratios are recovered within 15\% of the \hitomi\ SXS value, and within the uncertainty range of the SXS measurements; Ca/Fe and Cr/Fe are overestimated by 40\% (for Ca, this could be corrected by using \textit{only} the XIS result, which agrees with the SXS value), while Mn/Fe is overestimated by a large factor (200\%). `Local' fits help to reduce the Mn/Fe bias, but the measurement of this abundance ratio is still 80\% higher than the \hitomi\ measurement. `Local' and `global' averages are generally in agreement, although the `local' fits perform slightly worse for Ar/Fe and Ni/Fe and slightly better for Ca/Fe compared to `global' fits.

Hence, the systematic limit is quite quickly reached for abundance measurements using X-ray CCD spectroscopy of clusters. It is nevertheless quite remarkable that we are able to measure the abundance of most of the elements well within 40\% with a low-resolution X-ray spectrometer in an object at a distance of $\sim$70 Mpc, compared to methods in other wavebands. The SXS instrument aboard \hitomi\ has shown that high-resolution X-ray spectroscopy can help us reduce the systematic uncertainty budget substantially, and even improve the results from CCD spectrometers by making spectral models more accurate. In particular, the Ni/Fe measurements from CCD spectroscopy are now in better agreement with the SXS results largely because \hitomi\ allowed us to measure and update the emissivity of several Fe lines that appeared blended with Ni at lower spectral resolution.

\section{Conclusions}

We have combined the latest results obtained with the high spectral resolution X-ray micro-calorimeter onboard the Hitomi satellite, together with archival data, in order to perform an in-depth study of the chemical enrichment pattern in the diffuse intra-cluster medium at the core of the Perseus cluster of galaxies. 
Our results can be summarised as follows:
\begin{itemize}
\item taking both statistical and systematic uncertainties into account, the \hitomi\ SXS provides tight constraints on the abundances of 8 different chemical elements: Si, S, Ar, Ca, Cr, Mn, Fe, and Ni. The total uncertainty budget amounts to $\sim$10\% for the absolute Fe abundance as well as for S/Fe, Ar/Fe, Ca/Fe, Ni/Fe, $\sim$15\% for Si/Fe and Cr/Fe, and $\sim$20\% for Mn/Fe.
\item no lines from elements other than those listed in the previous point were detected with high confidence, although hints of Al Ly ${\beta}$, a blend of Cl Ly ${\alpha}$ and K He ${\alpha}$, and Ti He ${\alpha}$ are seen with a significance between $1-2\sigma$. Within the large statistical uncertainties, the abundances of these elements with respect to Fe are consistent with the Solar ratios.
\item while the absolute abundance (i.e., Fe/H) decreases by about 30\% with increasing radius, we find no significant spatial variations in the relative abundance ratios of other detected elements with respect to Fe over the area probed by the \hitomi\ SXS (extending from the AGN in the BCG out to about 100~kpc). 
\item emission lines from lighter elements like O, Ne, and Mg fall outside of the limited bandpass of the available SXS observations; currently, the best constraints on their relative abundances with respect to Fe are obtained from the \xmm\ RGS, with typical estimated uncertainties of $\sim$20--30\%.
\item both the RGS and SXS measurements of the chemical enrichment pattern in the core of the Perseus Cluster are remarkably consistent with the abundance ratios (relative to Fe) measured in the proto-solar nebula \citep{lodders2009}, in {\it low-mass} early type galaxies \citep{conroy2014}, and in optical and infrared studies of other typical Milky Way stars with near-solar absolute metallicity \citep{bensby2014,battistini2015,hawkins2016}. The simple assumption that all abundance ratios are equal to Solar gives an excellent description of the Perseus Cluster ICM, with $\chi^2=10.7$ for 10 d.o.f. and no free parameters. 
\end{itemize}

Furthermore, we have compared in detail the metal abundance ratios measured using high-resolution spectra (RGS and SXS) with those obtained from the \xmm\ EPIC and \suzaku\ XIS detectors for the same sky region, employing the latest updates to the atomic line emission data bases. We conclude that:
\begin{itemize}
\item there is a significant scatter among the results from CCD spectra; for each element, at least one choice of detector (XIS, MOS or pn) and fitting technique (`local' or `global', i.e. using narrow-band or broad-band energy ranges) matches the high-resolution spectroscopy result, but no single method and instrument stand out as being systematically more accurate over all. 
\item \textit{averaging} the broadband results from the three different CCD-type instruments considered here, we find that the Fe/H, Si/Fe, S/Fe, Ar/Fe, and Ni/Fe ratios for the core of the Perseus Cluster are recovered within the uncertainty range of the SXS measurements, while Ca/Fe and Cr/Fe are biased high by $\sim40$\% in the lower resolution spectra (with the Ca bias driven primarily by \xmm, while XIS and SXS values agree). The largest bias is found for Mn/Fe, where the average value derived from CCD spectroscopy is larger by a factor of 2--3 compared to the \hitomi\ measurement. 
\end{itemize}
We caution the reader, however, that a conclusive study of the level of systematic errors associated with abundance ratio measurements from X-ray CCDs would require high-spectral resolution observations of a larger sample of clusters spanning a range of average temperatures and redshifts.

As the ICM in clusters of galaxies incorporates the accumulated ejecta from billions of type Ia and core-collapse supernova explosions, the exact metal abundance pattern can be used to constrain the ratio of SN~Ia to SNcc contributing to the enrichment, as well as the exact details of the supernova explosion mechanisms and progenitors. Previous studies in this respect focused primarily on \xmm\ (and, to a lesser extent, \suzaku) measurements, which suffer from the systematic uncertainties discussed above. We have conducted a detailed analysis to assess how well current supernova yield calculations are able to reproduce the abundance pattern observed in the ICM, based on the most reliable high-resolution spectroscopy constraints available for the core of the Perseus Cluster. We find that:
\begin{itemize}
\item the `classical' SN~Ia yields of \citet{iwamoto1999} have been updated recently by \citet{nomoto2018} to include the latest electron capture and nuclear reaction rates, resulting in a substantially different Ni/Fe production. Contrary to arguments published in previous literature \citep[e.g.][]{dupke2000}, slow deflagration models for SN~Ia do not necessarily imply a highly super-solar Ni/Fe in the ICM; the latest W7 model is currently preferred over the updated 1D delayed detonation models, since the latter heavily overproduce Cr/Fe compared to the observations. Nevertheless, none of the `classical' SN~Ia yields can reproduce the relative abundances of Fe-group elements satisfactorily. 
\item mixing the \citet{nomoto2006} SNcc yields for a Solar progenitor metallicity with the \citet{leung2017} 2D delayed detonation near-$M_{Ch}$ SN~Ia model for zero metal abundance of the progenitor yields a formally acceptable fit, although requiring such a large discrepancy between the SNcc and SNIa progenitor compositions seems unphysical.
\item assuming the SNcc and SNIa progenitors to have the same initial metal abundance, the model that best reproduces both the light $\alpha$-element (O, Ne, Mg) and Fe-group element (Cr, Mn, Fe) ratios with respect to Fe is given by combining the SNcc yields of \citet{nomoto2006} together with the N100 delayed detonation plus the `1.1-0.9' violent merger for the near- and sub-$M_{Ch}$ explosion channels, respectively \citep{seitenzahl2013,pakmor2012}. This model suggests that SN~Ia represent $25\pm6$\% of the total number of supernovae, and near-$M_{Ch}$ SN~Ia represent $30\pm14$\% of the total number of SN~Ia. However, this model does not reproduce the relative abundances of intermediate $\alpha$-elements, in particular over-predicting the Si/Ar and Si/Ca ratios. 
\item a mixture of the `N20' SNcc model of \citet{sukhbold2016}, the N100 model for near-$M_{Ch}$ SN~Ia, and a recently published `dynamically-driven double-degenerate double-detonation' scenario for sub-$M_{Ch}$ SN~Ia \citep{shen2018} provides a better agreement with the measured Si/Ar and Si/Ca ratio, although S/Si is overproduced. This model suggests a SN~Ia fraction of $38\pm6$\%, with up to 20\% of SN~Ia progenitors being near-$M_{Ch}$. Due to the lower best-fit fraction of near-$M_{Ch}$ SN~Ia, this model on the other hand underproduces Mn/Fe; hence, currently, a better fit to the $\alpha$-element to Fe ratios comes at the expense of obtaining a worse description of the Fe-group to Fe composition.
\item the \hitomi\ SXS provides much tighter constraints on the abundance of Ar than stellar spectra or even measurements of the proto-solar composition. The precise measurements of Si/Ar and S/Ar enabled by \hitomi\ can be used to refine future core-collapse supernova yields. In particular, it seems that including the effects of neutrino transport in the SNcc yield calculation of \citet{sukhbold2016} can, in some model variants, reproduce the relative abundances of various intermediate $\alpha$-elements better than the \citet{nomoto2006} yields that do not account for these effects. 
\end{itemize}

Our results demonstrate the power of high-resolution X-ray spectroscopy in determining the origin of various chemical elements in the Universe. Future progress in this field will rely both on obtaining precise measurements of the metal abundance pattern in a larger sample of clusters of galaxies using high-resolution spectrometers on upcoming X-ray missions such as \textit{XRISM}-\textit{Resolve} \citep{XRISM} and \textit{Athena} X-IFU \citep{XIFU}, but also on improvements in our theoretical models of supernova nucleosynthesis.

\section*{Acknowledgments}

We thank all the members who have contributed to the ASTRO-H (\hitomi) team, and are grateful to T. Sukhbold for providing some of the supernova yield calculations used throughout this work, and to S. Beaumont and A. Boldog for their valuable input to this manuscript. We thank the anonymous referee for their comments and suggestions. A. Simionescu is supported by the Women In Science Excel (WISE) programme of the Netherlands Organisation for Scientific Research (NWO), and acknowledges the World Premier Research Center Initiative (WPI) and the Kavli IPMU for the continued hospitality. SRON Netherlands Institute for Space Research is supported financially by NWO.
S. Nakashima and L. Gu are supported by the RIKEN Special Postdoctoral Researcher Program. 
K. Matsushita acknowledges support from JSPS KAKENHI Grant Number JP16K05300.
F. Mernier and N. Werner are supported by the Lend\"ulet LP2016-11 grant awarded by the Hungarian Academy of Sciences.
K. Nomoto and S.-C. Leung acknowledge support from MEXT WPI and JSPS KAKENHI Grant Numbers JP26400222, JP16H02168 and JP17K05382. 
E. Bulbul and E. Miller acknowledge support from NASA grant NNX15AC76G. 
A.C. Fabian and C. Pinto acknowledge ERC Advanced Grant 340442.
Y. Ichinohe is supported by the Rikkyo University Special Fund for Research (SFR).
T. Kitayama acknowledges support from KAKENHI Grant Number 25400236. 
H. Noda is supported by the Program for Establishing a Consortium for the Development of Human Resources in Science and Technology, Japan Science and Technology Agency (JST).
S. Safi-Harb acknowledges support from NSERC and CSA. 
K. Sato is grateful for KAKENHI grant 17K05393.
S. Ueda is supported in part by the Ministry of Science and Technology of Taiwan (grant MOST 106-2628-M-001-003-MY3) and by Academia Sinica (grant AS-IA-107-M01).

\bibliographystyle{mnras}
\bibliography{clusters.bib}

\begin{thebibliography}{}
\makeatletter
\relax
\def\mn@urlcharsother{\let\do\@makeother \do\$\do\&\do\#\do\^\do\_\do\%\do\~}
\def\mn@doi{\begingroup\mn@urlcharsother \@ifnextchar [ {\mn@doi@}
  {\mn@doi@[]}}
\def\mn@doi@[#1]#2{\def\@tempa{#1}\ifx\@tempa\@empty \href
  {http://dx.doi.org/#2} {doi:#2}\else \href {http://dx.doi.org/#2} {#1}\fi
  \endgroup}
\def\mn@eprint#1#2{\mn@eprint@#1:#2::\@nil}
\def\mn@eprint@arXiv#1{\href {http://arxiv.org/abs/#1} {{\tt arXiv:#1}}}
\def\mn@eprint@dblp#1{\href {http://dblp.uni-trier.de/rec/bibtex/#1.xml}
  {dblp:#1}}
\def\mn@eprint@#1:#2:#3:#4\@nil{\def\@tempa {#1}\def\@tempb {#2}\def\@tempc
  {#3}\ifx \@tempc \@empty \let \@tempc \@tempb \let \@tempb \@tempa \fi \ifx
  \@tempb \@empty \def\@tempb {arXiv}\fi \@ifundefined
  {mn@eprint@\@tempb}{\@tempb:\@tempc}{\expandafter \expandafter \csname
  mn@eprint@\@tempb\endcsname \expandafter{\@tempc}}}

\bibitem[\protect\citeauthoryear{{Aharonian} et~al.,}{{Aharonian}
  et~al.}{2017}]{Hitomi35keV}
{Aharonian} F.~A.,  et~al., 2017, \mn@doi [\apjl] {10.3847/2041-8213/aa61fa},
  \href {http://adsabs.harvard.edu/abs/2017ApJ...837L..15A} {837, L15}

\bibitem[\protect\citeauthoryear{{Angelini} et~al.,}{{Angelini}
  et~al.}{2016}]{angelini2016}
{Angelini} L.,  et~al., 2016, in Space Telescopes and Instrumentation 2016:
  Ultraviolet to Gamma Ray. p. 990514, \mn@doi{10.1117/12.2234429}

\bibitem[\protect\citeauthoryear{{Arnaud}}{{Arnaud}}{1996}]{arnaud1996}
{Arnaud} K.~A.,  1996, in {Jacoby} G.~H.,  {Barnes} J.,  eds,  Astronomical
  Society of the Pacific Conference Series Vol. 101, Astronomical Data Analysis
  Software and Systems V. p.~17

\bibitem[\protect\citeauthoryear{{Asplund}, {Grevesse}, {Sauval}  \&
  {Scott}}{{Asplund} et~al.}{2009}]{asplund2009}
{Asplund} M.,  {Grevesse} N.,  {Sauval} A.~J.,   {Scott} P.,  2009, \mn@doi
  [\araa] {10.1146/annurev.astro.46.060407.145222}, \href
  {http://adsabs.harvard.edu/abs/2009ARA%26A..47..481A} {47, 481}

\bibitem[\protect\citeauthoryear{{Barret} et~al.,}{{Barret}
  et~al.}{2018}]{XIFU}
{Barret} D.,  et~al., 2018, in Society of Photo-Optical Instrumentation
  Engineers (SPIE) Conference Series. p. 106991G (\mn@eprint {arXiv}
  {1807.06092}), \mn@doi{10.1117/12.2312409}

\bibitem[\protect\citeauthoryear{{Battistini} \& {Bensby}}{{Battistini} \&
  {Bensby}}{2015}]{battistini2015}
{Battistini} C.,  {Bensby} T.,  2015, \mn@doi [\aap]
  {10.1051/0004-6361/201425327}, \href
  {http://adsabs.harvard.edu/abs/2015A%26A...577A...9B} {577, A9}

\bibitem[\protect\citeauthoryear{{Bensby}, {Feltzing}  \& {Oey}}{{Bensby}
  et~al.}{2014}]{bensby2014}
{Bensby} T.,  {Feltzing} S.,   {Oey} M.~S.,  2014, \mn@doi [\aap]
  {10.1051/0004-6361/201322631}, \href
  {http://adsabs.harvard.edu/abs/2014A%26A...562A..71B} {562, A71}

\bibitem[\protect\citeauthoryear{{Bulbul}, {Smith}  \& {Loewenstein}}{{Bulbul}
  et~al.}{2012}]{bulbul2012}
{Bulbul} E.,  {Smith} R.~K.,   {Loewenstein} M.,  2012, \mn@doi [\apj]
  {10.1088/0004-637X/753/1/54}, \href
  {http://adsabs.harvard.edu/abs/2012ApJ...753...54B} {753, 54}

\bibitem[\protect\citeauthoryear{{Burbidge}, {Burbidge}, {Fowler}  \&
  {Hoyle}}{{Burbidge} et~al.}{1957}]{burbidge1957}
{Burbidge} E.~M.,  {Burbidge} G.~R.,  {Fowler} W.~A.,   {Hoyle} F.,  1957,
  \mn@doi [Reviews of Modern Physics] {10.1103/RevModPhys.29.547}, \href
  {http://adsabs.harvard.edu/abs/1957RvMP...29..547B} {29, 547}

\bibitem[\protect\citeauthoryear{{Cash}}{{Cash}}{1979}]{cash1979}
{Cash} W.,  1979, \mn@doi [\apj] {10.1086/156922}, \href
  {http://adsabs.harvard.edu/abs/1979ApJ...228..939C} {228, 939}

\bibitem[\protect\citeauthoryear{{Conroy}, {Graves}  \& {van Dokkum}}{{Conroy}
  et~al.}{2014}]{conroy2014}
{Conroy} C.,  {Graves} G.~J.,   {van Dokkum} P.~G.,  2014, \mn@doi [\apj]
  {10.1088/0004-637X/780/1/33}, \href
  {http://adsabs.harvard.edu/abs/2014ApJ...780...33C} {780, 33}

\bibitem[\protect\citeauthoryear{{Conselice}, {Gallagher}  \&
  {Wyse}}{{Conselice} et~al.}{2001}]{conselice2001}
{Conselice} C.~J.,  {Gallagher} III J.~S.,   {Wyse} R.~F.~G.,  2001, \mn@doi
  [\aj] {10.1086/323534}, \href
  {http://cdsads.u-strasbg.fr/abs/2001AJ....122.2281C} {122, 2281}

\bibitem[\protect\citeauthoryear{{Dave}, {Kashyap}, {Fisher}, {Timmes},
  {Townsley}  \& {Byrohl}}{{Dave} et~al.}{2017}]{dave2017}
{Dave} P.,  {Kashyap} R.,  {Fisher} R.,  {Timmes} F.,  {Townsley} D.,
  {Byrohl} C.,  2017, \mn@doi [\apj] {10.3847/1538-4357/aa7134}, \href
  {http://adsabs.harvard.edu/abs/2017ApJ...841...58D} {841, 58}

\bibitem[\protect\citeauthoryear{{Dupke} \& {White}}{{Dupke} \&
  {White}}{2000}]{dupke2000}
{Dupke} R.~A.,  {White} III R.~E.,  2000, \mn@doi [\apj] {10.1086/308181},
  \href
  {http://adsabs.harvard.edu/cgi-bin/nph-bib_query?bibcode=2000ApJ...528..139D&db_key=AST}
  {528, 139}

\bibitem[\protect\citeauthoryear{{Ettori} \& {Fabian}}{{Ettori} \&
  {Fabian}}{1999}]{ettori1999}
{Ettori} S.,  {Fabian} A.~C.,  1999, \mnras, \href
  {http://adsabs.harvard.edu/abs/1999MNRAS.305..834E} {305, 834}

\bibitem[\protect\citeauthoryear{{Finoguenov}, {David}  \&
  {Ponman}}{{Finoguenov} et~al.}{2000}]{finoguenov2000}
{Finoguenov} A.,  {David} L.~P.,   {Ponman} T.~J.,  2000, \mn@doi [\apj]
  {10.1086/317173}, \href {http://adsabs.harvard.edu/abs/2000ApJ...544..188F}
  {544, 188}

\bibitem[\protect\citeauthoryear{{Foster}, {Ji}, {Smith}  \&
  {Brickhouse}}{{Foster} et~al.}{2012}]{foster2012}
{Foster} A.~R.,  {Ji} L.,  {Smith} R.~K.,   {Brickhouse} N.~S.,  2012, \mn@doi
  [\apj] {10.1088/0004-637X/756/2/128}, \href
  {http://adsabs.harvard.edu/abs/2012ApJ...756..128F} {756, 128}

\bibitem[\protect\citeauthoryear{{Fukazawa}, {Makishima}, {Tamura}, {Nakazawa},
  {Ezawa}, {Ikebe}, {Kikuchi}  \& {Ohashi}}{{Fukazawa}
  et~al.}{2000}]{fukazawa2000}
{Fukazawa} Y.,  {Makishima} K.,  {Tamura} T.,  {Nakazawa} K.,  {Ezawa} H.,
  {Ikebe} Y.,  {Kikuchi} K.,   {Ohashi} T.,  2000, \mn@doi [\mnras]
  {10.1046/j.1365-8711.2000.03204.x}, \href
  {http://adsabs.harvard.edu/abs/2000MNRAS.313...21F} {313, 21}

\bibitem[\protect\citeauthoryear{{Graham} et~al.,}{{Graham}
  et~al.}{2012}]{graham2012}
{Graham} M.~L.,  et~al., 2012, \mn@doi [\apj] {10.1088/0004-637X/753/1/68},
  \href {http://adsabs.harvard.edu/abs/2012ApJ...753...68G} {753, 68}

\bibitem[\protect\citeauthoryear{{Gu}, {Kaastra}  \& {Raassen}}{{Gu}
  et~al.}{2016}]{gu2016}
{Gu} L.,  {Kaastra} J.,   {Raassen} A.~J.~J.,  2016, \mn@doi [\aap]
  {10.1051/0004-6361/201527615}, \href
  {http://adsabs.harvard.edu/abs/2016A%26A...588A..52G} {588, A52}

\bibitem[\protect\citeauthoryear{{Hawkins}, {Masseron}, {Jofr{\'e}}, {Gilmore},
  {Elsworth}  \& {Hekker}}{{Hawkins} et~al.}{2016}]{hawkins2016}
{Hawkins} K.,  {Masseron} T.,  {Jofr{\'e}} P.,  {Gilmore} G.,  {Elsworth} Y.,
  {Hekker} S.,  2016, \mn@doi [\aap] {10.1051/0004-6361/201628812}, \href
  {http://adsabs.harvard.edu/abs/2016A%26A...594A..43H} {594, A43}

\bibitem[\protect\citeauthoryear{{Hitomi Collaboration}}{{Hitomi
  Collaboration}}{2017}]{HitomiNatureZ}
{Hitomi Collaboration} 2017, \mn@doi [\nat] {10.1038/nature24301}, \href
  {http://adsabs.harvard.edu/abs/2017Natur.551..478A} {551, 478}

\bibitem[\protect\citeauthoryear{{Hitomi Collaboration}}{{Hitomi
  Collaboration}}{2018a}]{HitomiV}
{Hitomi Collaboration} 2018a, \mn@doi [\pasj] {10.1093/pasj/psx138}, \href
  {http://adsabs.harvard.edu/abs/2018PASJ...70....9H} {70, 9}

\bibitem[\protect\citeauthoryear{{Hitomi Collaboration}}{{Hitomi
  Collaboration}}{2018b}]{HitomiRS}
{Hitomi Collaboration} 2018b, \mn@doi [\pasj] {10.1093/pasj/psx127}, \href
  {http://adsabs.harvard.edu/abs/2018PASJ...70...10H} {70, 10}

\bibitem[\protect\citeauthoryear{{Hitomi Collaboration}}{{Hitomi
  Collaboration}}{2018c}]{HitomiT}
{Hitomi Collaboration} 2018c, \mn@doi [\pasj] {10.1093/pasj/psy004}, \href
  {http://adsabs.harvard.edu/abs/2018PASJ...70...11H} {70, 11}

\bibitem[\protect\citeauthoryear{{Hitomi Collaboration}}{{Hitomi
  Collaboration}}{2018d}]{HitomiAtomic}
{Hitomi Collaboration} 2018d, \mn@doi [\pasj] {10.1093/pasj/psx156}, \href
  {http://adsabs.harvard.edu/abs/2018PASJ...70...12H} {70, 12}

\bibitem[\protect\citeauthoryear{{Hitomi Collaboration}}{{Hitomi
  Collaboration}}{2018e}]{HitomiAGN}
{Hitomi Collaboration} 2018e, \mn@doi [\pasj] {10.1093/pasj/psx147}, \href
  {http://adsabs.harvard.edu/abs/2018PASJ...70...13H} {70, 13}

\bibitem[\protect\citeauthoryear{{Iwamoto}, {Brachwitz}, {Nomoto}, {Kishimoto},
  {Umeda}, {Hix}  \& {Thielemann}}{{Iwamoto} et~al.}{1999}]{iwamoto1999}
{Iwamoto} K.,  {Brachwitz} F.,  {Nomoto} K.,  {Kishimoto} N.,  {Umeda} H.,
  {Hix} W.~R.,   {Thielemann} F.,  1999, \apjs, 125, 439

\bibitem[\protect\citeauthoryear{{Jacobson} et~al.,}{{Jacobson}
  et~al.}{2015}]{jacobson2015}
{Jacobson} H.~R.,  et~al., 2015, \mn@doi [\apj] {10.1088/0004-637X/807/2/171},
  \href {http://adsabs.harvard.edu/abs/2015ApJ...807..171J} {807, 171}

\bibitem[\protect\citeauthoryear{{Jha}}{{Jha}}{2017}]{jha2017}
{Jha} S.~W.,  2017, in Handbook of Supernovae, ed. A.~W. Alsabti \& P.~Murdin (Springer), 1, 375 (\mn@eprint {arXiv}{1707.01110}).

\bibitem[\protect\citeauthoryear{{Kaastra}, {Mewe}  \&
  {Nieuwenhuijzen}}{{Kaastra} et~al.}{1996}]{kaastra1996}
{Kaastra} J.~S.,  {Mewe} R.,   {Nieuwenhuijzen} H.,  1996, in UV and X-ray
  Spectroscopy of Astrophysical and Laboratory Plasmas p.411, K. Yamashita and
  T. Watanabe. Tokyo : Universal Academy Press.

\bibitem[\protect\citeauthoryear{{Kalberla}, {Burton}, {Hartmann}, {Arnal},
  {Bajaja}, {Morras}  \& {P{\"o}ppel}}{{Kalberla} et~al.}{2005}]{kalberla2005}
{Kalberla} P.~M.~W.,  {Burton} W.~B.,  {Hartmann} D.,  {Arnal} E.~M.,  {Bajaja}
  E.,  {Morras} R.,   {P{\"o}ppel} W.~G.~L.,  2005, \mn@doi [\aap]
  {10.1051/0004-6361:20041864}, \href
  {http://adsabs.harvard.edu/abs/2005A%26A...440..775K} {440, 775}

\bibitem[\protect\citeauthoryear{{Kitayama} et~al.,}{{Kitayama}
  et~al.}{2014}]{kitayama2014}
{Kitayama} T.,  et~al., 2014, preprint, \href
  {http://adsabs.harvard.edu/abs/2014arXiv1412.1176K} {} (\mn@eprint {arXiv}
  {1412.1176})

\bibitem[\protect\citeauthoryear{{Kobayashi}, {Umeda}, {Nomoto}, {Tominaga}  \&
  {Ohkubo}}{{Kobayashi} et~al.}{2006}]{kobayashi2006}
{Kobayashi} C.,  {Umeda} H.,  {Nomoto} K.,  {Tominaga} N.,   {Ohkubo} T.,
  2006, \mn@doi [\apj] {10.1086/508914}, \href
  {http://adsabs.harvard.edu/abs/2006ApJ...653.1145K} {653, 1145}

\bibitem[\protect\citeauthoryear{{Leung} \& {Nomoto}}{{Leung} \&
  {Nomoto}}{2018}]{leung2017}
{Leung} S.-C.,  {Nomoto} K.,  2018, \mn@doi [\apj] {10.3847/1538-4357/aac2df},
  \href {http://adsabs.harvard.edu/abs/2018ApJ...861..143L} {861, 143}

\bibitem[\protect\citeauthoryear{{Lodders}, {Palme}  \& {Gail}}{{Lodders}
  et~al.}{2009}]{lodders2009}
{Lodders} K.,  {Palme} H.,   {Gail} H.-P.,  2009, \mn@doi [Landolt
  B{\"o}rnstein] {10.1007/978-3-540-88055-4_34}, \href
  {http://adsabs.harvard.edu/abs/2009LanB...4B...44L} {}

\bibitem[\protect\citeauthoryear{{Mannucci}, {Maoz}, {Sharon}, {Botticella},
  {Della Valle}, {Gal-Yam}  \& {Panagia}}{{Mannucci}
  et~al.}{2008}]{mannucci2008}
{Mannucci} F.,  {Maoz} D.,  {Sharon} K.,  {Botticella} M.~T.,  {Della Valle}
  M.,  {Gal-Yam} A.,   {Panagia} N.,  2008, \mn@doi [\mnras]
  {10.1111/j.1365-2966.2007.12603.x}, \href
  {http://adsabs.harvard.edu/abs/2008MNRAS.383.1121M} {383, 1121}

\bibitem[\protect\citeauthoryear{{Mantz}, {Allen}, {Morris}, {Simionescu},
  {Urban}, {Werner}  \& {Zhuravleva}}{{Mantz} et~al.}{2017}]{mantz2017}
{Mantz} A.~B.,  {Allen} S.~W.,  {Morris} R.~G.,  {Simionescu} A.,  {Urban} O.,
  {Werner} N.,   {Zhuravleva} I.,  2017, \mn@doi [\mnras]
  {10.1093/mnras/stx2200}, \href
  {http://adsabs.harvard.edu/abs/2017MNRAS.472.2877M} {472, 2877}

\bibitem[\protect\citeauthoryear{{Maoz} \& {Graur}}{{Maoz} \&
  {Graur}}{2017}]{maoz2017}
{Maoz} D.,  {Graur} O.,  2017, \mn@doi [\apj] {10.3847/1538-4357/aa8b6e}, \href
  {http://adsabs.harvard.edu/abs/2017ApJ...848...25M} {848, 25}

\bibitem[\protect\citeauthoryear{{Matsushita}, {B{\"o}hringer}, {Takahashi}  \&
  {Ikebe}}{{Matsushita} et~al.}{2007}]{matsushita2007}
{Matsushita} K.,  {B{\"o}hringer} H.,  {Takahashi} I.,   {Ikebe} Y.,  2007,
  \mn@doi [\aap] {10.1051/0004-6361:20041577}, \href
  {http://adsabs.harvard.edu/cgi-bin/nph-bib_query?bibcode=2007A%26A...462..953M&db_key=AST}
  {462, 953}

\bibitem[\protect\citeauthoryear{{Matsushita}, {Sakuma}, {Sasaki}, {Sato}  \&
  {Simionescu}}{{Matsushita} et~al.}{2013}]{matsushita2013}
{Matsushita} K.,  {Sakuma} E.,  {Sasaki} T.,  {Sato} K.,   {Simionescu} A.,
  2013, \mn@doi [\apj] {10.1088/0004-637X/764/2/147}, \href
  {http://adsabs.harvard.edu/abs/2013ApJ...764..147M} {764, 147}

\bibitem[\protect\citeauthoryear{{Mernier}, {de Plaa}, {Lovisari}, {Pinto},
  {Zhang}, {Kaastra}, {Werner}  \& {Simionescu}}{{Mernier}
  et~al.}{2015}]{mernier2015}
{Mernier} F.,  {de Plaa} J.,  {Lovisari} L.,  {Pinto} C.,  {Zhang} Y.-Y.,
  {Kaastra} J.~S.,  {Werner} N.,   {Simionescu} A.,  2015, \mn@doi [\aap]
  {10.1051/0004-6361/201425282}, \href
  {http://adsabs.harvard.edu/abs/2015A%26A...575A..37M} {575, A37}

\bibitem[\protect\citeauthoryear{{Mernier}, {de Plaa}, {Pinto}, {Kaastra},
  {Kosec}, {Zhang}, {Mao}  \& {Werner}}{{Mernier} et~al.}{2016a}]{mernier2016i}
{Mernier} F.,  {de Plaa} J.,  {Pinto} C.,  {Kaastra} J.~S.,  {Kosec} P.,
  {Zhang} Y.-Y.,  {Mao} J.,   {Werner} N.,  2016a, \mn@doi [\aap]
  {10.1051/0004-6361/201527824}, \href
  {http://adsabs.harvard.edu/abs/2016A%26A...592A.157M} {592, A157}

\bibitem[\protect\citeauthoryear{{Mernier} et~al.,}{{Mernier}
  et~al.}{2016b}]{mernier2016ii}
{Mernier} F.,  et~al., 2016b, \mn@doi [\aap] {10.1051/0004-6361/201628765},
  \href {http://adsabs.harvard.edu/abs/2016A%26A...595A.126M} {595, A126}

\bibitem[\protect\citeauthoryear{{Mernier} et~al.,}{{Mernier}
  et~al.}{2017}]{mernier2017}
{Mernier} F.,  et~al., 2017, \mn@doi [\aap] {10.1051/0004-6361/201630075},
  \href {http://adsabs.harvard.edu/abs/2017A%26A...603A..80M} {603, A80}

\bibitem[\protect\citeauthoryear{{Mernier} et~al.,}{{Mernier}
  et~al.}{2018}]{mernier_rev}
{Mernier} F.,  et~al., 2018, preprint, \href
  {http://esoads.eso.org/abs/2018arXiv181101967M} {} (\mn@eprint {arXiv}
  {1811.01967})

\bibitem[\protect\citeauthoryear{{Mulchaey}, {Kasliwal}  \&
  {Kollmeier}}{{Mulchaey} et~al.}{2014}]{mulchaey2014}
{Mulchaey} J.~S.,  {Kasliwal} M.~M.,   {Kollmeier} J.~A.,  2014, \mn@doi
  [\apjl] {10.1088/2041-8205/780/2/L34}, \href
  {http://adsabs.harvard.edu/abs/2014ApJ...780L..34M} {780, L34}

\bibitem[\protect\citeauthoryear{{Mushotzky}, {Loewenstein}, {Arnaud},
  {Tamura}, {Fukazawa}, {Matsushita}, {Kikuchi}  \& {Hatsukade}}{{Mushotzky}
  et~al.}{1996}]{mushotzky1996}
{Mushotzky} R.,  {Loewenstein} M.,  {Arnaud} K.~A.,  {Tamura} T.,  {Fukazawa}
  Y.,  {Matsushita} K.,  {Kikuchi} K.,   {Hatsukade} I.,  1996, \mn@doi [\apj]
  {10.1086/177541}, \href
  {http://adsabs.harvard.edu/cgi-bin/nph-bib_query?bibcode=1996ApJ...466..686M&db_key=AST}
  {466, 686}

\bibitem[\protect\citeauthoryear{{Nomoto} \& {Hashimoto}}{{Nomoto} \&
  {Hashimoto}}{1988}]{nomoto1988}
{Nomoto} K.,  {Hashimoto} M.,  1988, \mn@doi [\physrep]
  {10.1016/0370-1573(88)90032-4}, 163, 13

\bibitem[\protect\citeauthoryear{{Nomoto} \& {Leung}}{{Nomoto} \&
  {Leung}}{2018}]{nomoto2018}
{Nomoto} K.,  {Leung} S.-C.,  2018, \mn@doi [\ssr] {10.1007/s11214-018-0499-0},
  \href {http://adsabs.harvard.edu/abs/2018SSRv..214...67N} {214, 67}

\bibitem[\protect\citeauthoryear{{Nomoto}, {Tominaga}, {Umeda}, {Kobayashi}  \&
  {Maeda}}{{Nomoto} et~al.}{2006}]{nomoto2006}
{Nomoto} K.,  {Tominaga} N.,  {Umeda} H.,  {Kobayashi} C.,   {Maeda} K.,  2006,
  \mn@doi [Nuclear Physics A] {10.1016/j.nuclphysa.2006.05.008}, \href
  {http://adsabs.harvard.edu/abs/2006NuPhA.777..424N} {777, 424}

\bibitem[\protect\citeauthoryear{{Nomoto}, {Kobayashi}  \& {Tominaga}}{{Nomoto}
  et~al.}{2013}]{nomoto2013}
{Nomoto} K.,  {Kobayashi} C.,   {Tominaga} N.,  2013, \mn@doi [\araa]
  {10.1146/annurev-astro-082812-140956}, \href
  {http://supernova.astron.s.u-tokyo.ac.jp/\~nomoto/reference} {51, 457}

\bibitem[\protect\citeauthoryear{{Pakmor}, {Kromer}, {Taubenberger}, {Sim},
  {R{\"o}pke}  \& {Hillebrandt}}{{Pakmor} et~al.}{2012}]{pakmor2012}
{Pakmor} R.,  {Kromer} M.,  {Taubenberger} S.,  {Sim} S.~A.,  {R{\"o}pke}
  F.~K.,   {Hillebrandt} W.,  2012, \mn@doi [\apjl]
  {10.1088/2041-8205/747/1/L10}, \href
  {http://adsabs.harvard.edu/abs/2012ApJ...747L..10P} {747, L10}

\bibitem[\protect\citeauthoryear{{Rasmussen} \& {Ponman}}{{Rasmussen} \&
  {Ponman}}{2009}]{rasmussen2009}
{Rasmussen} J.,  {Ponman} T.~J.,  2009, \mn@doi [\mnras]
  {10.1111/j.1365-2966.2009.15244.x}, \href
  {http://adsabs.harvard.edu/abs/2009MNRAS.399..239R} {399, 239}

\bibitem[\protect\citeauthoryear{{Reggiani}, {Mel{\'e}ndez}, {Kobayashi},
  {Karakas}  \& {Placco}}{{Reggiani} et~al.}{2017}]{reggiani2017}
{Reggiani} H.,  {Mel{\'e}ndez} J.,  {Kobayashi} C.,  {Karakas} A.,   {Placco}
  V.,  2017, \mn@doi [\aap] {10.1051/0004-6361/201730750}, \href
  {http://adsabs.harvard.edu/abs/2017A%26A...608A..46R} {608, A46}

\bibitem[\protect\citeauthoryear{{Saio}, {Nomoto}  \& {Kato}}{{Saio}
  et~al.}{1988}]{saio1988}
{Saio} H.,  {Nomoto} K.,   {Kato} M.,  1988, \mn@doi [\nat] {10.1038/334508a0},
  344, 508

\bibitem[\protect\citeauthoryear{{Salpeter}}{{Salpeter}}{1955}]{salpeter1955}
{Salpeter} E.~E.,  1955, \mn@doi [\apj] {10.1086/145971}, \href
  {http://adsabs.harvard.edu/abs/1955ApJ...121..161S} {121, 161}

\bibitem[\protect\citeauthoryear{{Sanders} \& {Fabian}}{{Sanders} \&
  {Fabian}}{2006}]{sanders2006}
{Sanders} J.~S.,  {Fabian} A.~C.,  2006, \mn@doi [\mnras]
  {10.1111/j.1365-2966.2006.10779.x}, \href
  {http://adsabs.harvard.edu/abs/2006MNRAS.371.1483S} {371, 1483}

\bibitem[\protect\citeauthoryear{{Sato}, {Tokoi}, {Matsushita}, {Ishisaki},
  {Yamasaki}, {Ishida}  \& {Ohashi}}{{Sato} et~al.}{2007}]{sato2007}
{Sato} K.,  {Tokoi} K.,  {Matsushita} K.,  {Ishisaki} Y.,  {Yamasaki} N.~Y.,
  {Ishida} M.,   {Ohashi} T.,  2007, \mn@doi [\apjl] {10.1086/522031}, \href
  {http://adsabs.harvard.edu/abs/2007ApJ...667L..41S} {667, L41}

\bibitem[\protect\citeauthoryear{{Seitenzahl} et~al.,}{{Seitenzahl}
  et~al.}{2013}]{seitenzahl2013}
{Seitenzahl} I.~R.,  et~al., 2013, \mn@doi [\mnras] {10.1093/mnras/sts402},
  \href {http://adsabs.harvard.edu/abs/2013MNRAS.429.1156S} {429, 1156}

\bibitem[\protect\citeauthoryear{{Shen}, {Kasen}, {Miles}  \&
  {Townsley}}{{Shen} et~al.}{2018}]{shen2018}
{Shen} K.~J.,  {Kasen} D.,  {Miles} B.~J.,   {Townsley} D.~M.,  2018, \mn@doi
  [\apj] {10.3847/1538-4357/aaa8de}, \href
  {http://adsabs.harvard.edu/abs/2018ApJ...854...52S} {854, 52}

\bibitem[\protect\citeauthoryear{{Shigeyama} \& {Nomoto}}{{Shigeyama} \&
  {Nomoto}}{1990}]{shigeyama1990}
{Shigeyama} T.,  {Nomoto} K.,  1990, \mn@doi [\apj] {10.1086/169114}, \href
  {http://adsabs.harvard.edu/abs/1990ApJ...360..242S} {360, 242}

\bibitem[\protect\citeauthoryear{{Simionescu}, {Roediger}, {Nulsen},
  {Br{\"u}ggen}, {Forman}, {B{\"o}hringer}, {Werner}  \&
  {Finoguenov}}{{Simionescu} et~al.}{2009}]{simionescu2009a}
{Simionescu} A.,  {Roediger} E.,  {Nulsen} P.~E.~J.,  {Br{\"u}ggen} M.,
  {Forman} W.~R.,  {B{\"o}hringer} H.,  {Werner} N.,   {Finoguenov} A.,  2009,
  \mn@doi [\aap] {10.1051/0004-6361:200811071}, \href
  {http://adsabs.harvard.edu/abs/2009A%26A...495..721S} {495, 721}

\bibitem[\protect\citeauthoryear{{Simionescu}, {Werner}, {Urban}, {Allen},
  {Ichinohe}  \& {Zhuravleva}}{{Simionescu} et~al.}{2015}]{simionescu2015}
{Simionescu} A.,  {Werner} N.,  {Urban} O.,  {Allen} S.~W.,  {Ichinohe} Y.,
  {Zhuravleva} I.,  2015, \mn@doi [\apjl] {10.1088/2041-8205/811/2/L25}, \href
  {http://adsabs.harvard.edu/abs/2015ApJ...811L..25S} {811, L25}

\bibitem[\protect\citeauthoryear{{Sukhbold}, {Ertl}, {Woosley}, {Brown}  \&
  {Janka}}{{Sukhbold} et~al.}{2016}]{sukhbold2016}
{Sukhbold} T.,  {Ertl} T.,  {Woosley} S.~E.,  {Brown} J.~M.,   {Janka} H.-T.,
  2016, \mn@doi [\apj] {10.3847/0004-637X/821/1/38}, \href
  {http://adsabs.harvard.edu/abs/2016ApJ...821...38S} {821, 38}

\bibitem[\protect\citeauthoryear{{Takahashi} et~al.,}{{Takahashi}
  et~al.}{2016}]{takahashi2016}
{Takahashi} T.,  et~al., 2016, in Space Telescopes and Instrumentation 2016:
  Ultraviolet to Gamma Ray. p. 99050U, \mn@doi{10.1117/12.2232379}

\bibitem[\protect\citeauthoryear{{Tamura} et~al.,}{{Tamura}
  et~al.}{2009}]{tamura2009}
{Tamura} T.,  et~al., 2009, \mn@doi [\apjl] {10.1088/0004-637X/705/1/L62},
  \href {http://adsabs.harvard.edu/abs/2009ApJ...705L..62T} {705, L62}

\bibitem[\protect\citeauthoryear{{Tashiro} et~al.,}{{Tashiro}
  et~al.}{2018}]{XRISM}
{Tashiro} M.,  et~al., 2018, in Society of Photo-Optical Instrumentation
  Engineers (SPIE) Conference Series. p. 1069922, \mn@doi{10.1117/12.2309455}

\bibitem[\protect\citeauthoryear{{Tsujimoto}, {Nomoto}, {Yoshii}, {Hashimoto},
  {Yanagida}  \& {Thielemann}}{{Tsujimoto} et~al.}{1995}]{tsujimoto1995}
{Tsujimoto} T.,  {Nomoto} K.,  {Yoshii} Y.,  {Hashimoto} M.,  {Yanagida} S.,
  {Thielemann} F.-K.,  1995, \mnras, 277, 945

\bibitem[\protect\citeauthoryear{{Tsujimoto} et~al.,}{{Tsujimoto}
  et~al.}{2018}]{HitomiCrabARF}
{Tsujimoto} M.,  et~al., 2018, \mn@doi [\pasj] {10.1093/pasj/psy008}, \href
  {http://adsabs.harvard.edu/abs/2018PASJ...70...20T} {70, 20}

\bibitem[\protect\citeauthoryear{{Utrobin}, {Wongwathanarat}, {Janka}  \&
  {M{\"u}ller}}{{Utrobin} et~al.}{2015}]{utrobin2015}
{Utrobin} V.~P.,  {Wongwathanarat} A.,  {Janka} H.-T.,   {M{\"u}ller} E.,
  2015, \mn@doi [\aap] {10.1051/0004-6361/201425513}, \href
  {http://adsabs.harvard.edu/abs/2015A%26A...581A..40U} {581, A40}

\bibitem[\protect\citeauthoryear{{Waldman}, {Sauer}, {Livne}, {Perets},
  {Glasner}, {Mazzali}, {Truran}  \& {Gal-Yam}}{{Waldman}
  et~al.}{2011}]{waldman2011}
{Waldman} R.,  {Sauer} D.,  {Livne} E.,  {Perets} H.,  {Glasner} A.,  {Mazzali}
  P.,  {Truran} J.~W.,   {Gal-Yam} A.,  2011, \mn@doi [\apj]
  {10.1088/0004-637X/738/1/21}, \href
  {http://adsabs.harvard.edu/abs/2011ApJ...738...21W} {738, 21}

\bibitem[\protect\citeauthoryear{{Werner}, {Durret}, {Ohashi}, {Schindler}  \&
  {Wiersma}}{{Werner} et~al.}{2008}]{werner2008}
{Werner} N.,  {Durret} F.,  {Ohashi} T.,  {Schindler} S.,   {Wiersma} R.~P.~C.,
   2008, \mn@doi [\ssr] {10.1007/s11214-008-9320-9}, \href
  {http://adsabs.harvard.edu/abs/2008SSRv..134..337W} {134, 337}

\bibitem[\protect\citeauthoryear{{Werner}, {Urban}, {Simionescu}  \&
  {Allen}}{{Werner} et~al.}{2013}]{werner2013}
{Werner} N.,  {Urban} O.,  {Simionescu} A.,   {Allen} S.~W.,  2013, \mn@doi
  [\nat] {10.1038/nature12646}, \href
  {http://adsabs.harvard.edu/abs/2013Natur.502..656W} {502, 656}

\bibitem[\protect\citeauthoryear{{Wilms}, {Allen}  \& {McCray}}{{Wilms}
  et~al.}{2000}]{wilms2000}
{Wilms} J.,  {Allen} A.,   {McCray} R.,  2000, \apj, \href
  {http://adsabs.harvard.edu/cgi-bin/nph-bib_query?bibcode=2000ApJ...542..914W&db_key=AST}
  {542, 914}

\bibitem[\protect\citeauthoryear{{Yamaguchi} et~al.,}{{Yamaguchi}
  et~al.}{2015}]{yamaguchi2015}
{Yamaguchi} H.,  et~al., 2015, \mn@doi [\apjl] {10.1088/2041-8205/801/2/L31},
  \href {http://adsabs.harvard.edu/abs/2015ApJ...801L..31Y} {801, L31}

\bibitem[\protect\citeauthoryear{{de Plaa}, {Werner}, {Bleeker}, {Vink},
  {Kaastra}  \& {M{\'e}ndez}}{{de Plaa} et~al.}{2007}]{deplaa2007}
{de Plaa} J.,  {Werner} N.,  {Bleeker} J.~A.~M.,  {Vink} J.,  {Kaastra} J.~S.,
   {M{\'e}ndez} M.,  2007, \mn@doi [\aap] {10.1051/0004-6361:20066382}, \href
  {http://adsabs.harvard.edu/cgi-bin/nph-bib_query?bibcode=2007A%26A...465..345D&db_key=AST}
  {465, 345}

\bibitem[\protect\citeauthoryear{{de Plaa} et~al.,}{{de Plaa}
  et~al.}{2017}]{deplaa2017}
{de Plaa} J.,  et~al., 2017, \mn@doi [\aap] {10.1051/0004-6361/201629926},
  \href {http://adsabs.harvard.edu/abs/2017A%26A...607A..98D} {607, A98}

\makeatother
\end{thebibliography}

\end{document}